\documentclass[useAMS,usenatbib]{mn2e}
\usepackage[british]{babel}	
\usepackage{amsmath}
\usepackage[T1]{fontenc}
\usepackage{verbatim}
\usepackage{graphicx,subfigure,float}
\usepackage[font=small,hang]{caption}
\usepackage{txfonts}
\usepackage{upgreek}
\usepackage{hyperref}
\usepackage{tikz}
\usetikzlibrary{calc,trees,positioning,arrows,chains,shapes.geometric,%
    decorations.pathreplacing,decorations.pathmorphing,shapes,%
    matrix,shapes.symbols}

\def\hi{\ifmmode{\rm HI}\else{H\/{\sc i}}\fi} 
\def\siiii{\ifmmode{\rm SiIII}\else{Si\/{\sc iii}}\fi} 
\def\siii{\ifmmode{\rm SiII}\else{Si\/{\sc ii}}\fi} 
\def\sivi{\ifmmode{\rm SiVI}\else{Si\/{\sc vi}}\fi}
\def\cvi{\ifmmode{\rm CVI}\else{C\/{\sc vi}}\fi}
\def\cii{\ifmmode{\rm CII}\else{C\/{\sc ii}}\fi} 
\def\ciii{\ifmmode{\rm CIII}\else{C\/{\sc iii}}\fi} 
\def\mgii{\ifmmode{\rm MgII}\else{Mg\/{\sc ii}}\fi} 
\def\ovi{\ifmmode{\rm OVI}\else{O\/{\sc vi}}\fi}
\def\ovii{\ifmmode{\rm OVII}\else{O\/{\sc vii}}\fi}
\def\oviii{\ifmmode{\rm OVIII}\else{O\/{\sc viii}}\fi}

\newcommand {\kms} {\,{\rm km\,s}^{-1}}

\newcommand {\mo}{{\rm M}_\odot}

\newcommand {\ha}{{\rm H}\upalpha}
\newcommand {\Lya}{{\rm Ly}\upalpha}

\title[Gas clouds in the CGM of Milky Way-like galaxies]{The survival of gas clouds in the Circumgalactic Medium of Milky Way-like galaxies}
\author[L. Armillotta et al.]{L. Armillotta$^{1,2}$\thanks{E-mail: lucia.armillotta@anu.edu.au},  F. Fraternali$^{2,3}$, J. K. Werk$^{4}$, J. X. Prochaska$^{5}$, F. Marinacci$^{6}$\\
$^{1}$Research School of Astronomy and Astrophysics - The Australian National University, Canberra, ACT, 2611, Australia\\
$^{2}$Department of Physics and Astronomy, University of Bologna, Viale Berti Pichat 6/2, 40127 Bologna, Italy\\
$^{3}$Kapteyn Astronomical Institute, University of Groningen, 9747 AD Groningen, The Netherlands\\
$^{4}$University of Washington, Physics and Astronomy Building,  3910 15th Ave. NE, Room C319 Seattle, WA USA 98195-0002\\
$^{5}$UCO/Lick Observatory, University of California, Santa Cruz, CA, USA\\
$^{6}$Department of Physics, Kavli Institute for Astrophysics and Space Research, Massachusetts Institute of Technology, Cambridge, MA 02139, USA}

\begin{document}

\pagerange{\pageref{firstpage}--\pageref{lastpage}} \pubyear{2017}

\maketitle

\label{firstpage}

\begin{abstract}
Observational evidence shows that low-redshift galaxies are surrounded by extended haloes of multiphase gas, the so-called `circumgalactic medium' (CGM). To study the survival of relatively cool gas (T $< 10^5$ K) in the CGM, we performed a set of hydrodynamical simulations of cold (T $= 10^4$ K) neutral gas clouds travelling through a hot (T $=2\times10^6$ K) and low-density (n $=10^{-4}$ cm$^{-3}$) coronal medium, typical of Milky Way-like galaxies at large galactocentric distances ($\sim 50-150$ kpc). We explored the effects of different initial values of relative velocity and radius of the clouds. Our simulations were performed on a two-dimensional grid with constant mesh size (2 pc) and they include radiative cooling, photoionization heating and thermal conduction. We found that for large clouds (radii larger than 250 pc) the cool gas survives for very long time (larger than 250 Myr): despite that they are partially destroyed and fragmented into smaller cloudlets during their trajectory, the total mass of cool gas decreases at very low rates. We found that thermal conduction plays a significant role: its effect is
to hinder formation of hydrodynamical instabilities at the cloud-corona interface, keeping
the cloud compact and therefore more difficult to destroy.
The distribution of column densities extracted from our simulations are compatible with those observed for low-temperature ions (e.g. \siii\ and \siiii) and for high-temperature ions (\ovi) once we take into account that \ovi\ covers much more extended regions than the cool gas and, therefore, it is more likely to be detected along a generic line of sight. 

\end{abstract}

\begin{keywords}
conduction -- hydrodynamics -- methods: numerical -- galaxies: haloes -- galaxies: intergalactic medium
\end{keywords}

\section{Introduction}
\label{Introduction}
In the last years several pieces of evidence have shown the presence of multiphase gas in the halo regions of low-redshift galaxies. This multiphase halo gas, called the `circumgalactic medium' (CGM), is explained by the existence of flows of gas towards and away from galaxies, in agreement with the idea of a strong interplay between galaxies and their intergalactic environment. Thus, studying the nature of the CGM is essential to understand how galaxies evolve in their environments.

The presence of a diffuse and hot circumgalactic phase is predicted by current cosmological theories. Low-redshift galaxies are expected to be surrounded by gas atmospheres at the virial temperature ($T \gtrsim 10^6$ K for Milky Way galaxies), the so-called `cosmological coronae' \citep{Fukugita&Peebles06}, extending out to hundreds of kpc from the galaxy center. Unfortunately the X-ray surface brightness of these coronae is rather faint, limiting their possibility of detection \citep{Bregman07}. In the Milky Way, the existence of a hot corona was originally postulated by \citet{Spitzer56} as the medium that provides the pressure required to confine the High-Velocity Clouds \citep[HVCs, e.g.][]{Wakker&vanWoerden97}. Most of the evidence of a hot corona collected over the last decades has been indirect and comes, for instance, from the rotation measure of pulsars in the Large Magellanic Cloud \citep{Gaensler+08,Anderson&Bregman10}, \ovi\ absorption lines in spectra of halo stars or extragalactic sources \citep{Sembach+03, Fox+10}. Moreover,  the head-tail structure of almost all the HVCs \citep[][]{Bruens+00, Putman+11}, the asymmetry of the Magellanic Stream \citep[MS,][]{Mastropietro+05} and the loss of external gas from the dwarf spheroidal galaxies of the Local Group \citep{Grcevich&Putman09, Gatto+13, Salem+15}, may all be explained by ram-pressure exerted by the coronal medium. 
Recently, the presence of a hot corona around our Galaxy was revealed by combining detections of \ovii\ and \oviii\ absorption lines in quasar spectra and emission lines in the soft X-ray background \citep{Miller&Bregman15}. 

Haloes of hot gas have also been observed around some massive spiral galaxies. In these cases X-ray emission has been detected at more than 50 kpc from the center, indicating the presence of extended structures. Once extrapolated to the virial radius, the mass of these coronae is comparable with the mass of the baryonic discs of these galaxies ($\sim 10^{11}$ M$_\mathrm{\odot}$), accounting for $\sim 10-50\%$ of their associated `missing baryons' \citep[e.g.][]{Dai+12, Bogdan+13, Anderson+16}. By combining observations of \ovi\ absorbers around star-forming galaxies (from the COS-Halos survey, see below) together with the \ovii\ and \oviii\ absorption associated with our Galaxy in a single model of corona, \citet{Faerman+16} found that the typical coronal gas mass of a Milky-Way-like galaxy is $\sim 1.35\times 10^{11} \mo$.

Regarding the cooler phase of the CGM, \hi\ observations in the Milky Way revealed a population of HVCs, characterized by velocities inconsistent with the rotation of the Galactic disc \citep[e.g.][]{Wakker&vanWoerden97} and generally located at distances of $\sim 5-10$ kpc from the Galactic disc \citep[e.g][] {Wakker01,Wakker07}. The MS and its leading arm (LA) are an exception. They are situated in the interval at $\sim 50-200$ kpc from the disc \citep{Putman+12} and produced by interaction between the the Large Magellanic Cloud, the Small Magellanic Cloud and the Milky Way. Notably, the LA is composed by a number of head-tail compact HVCs larger than the MS, suggesting that they are caused by ram pressure stripping exerted by the coronal medium of the Milky Way \citep{Venzmer+12,For+13}.
The presence of the HVCs in the CGM around our Galaxy provides therefore direct evidence of a cold medium ($T \lesssim 10^4$ K). Furthermore, a significant fraction of these HVCs exhibits absorptions from elements at high ionization state \citep[e.g. \ovi, \cvi, \sivi,][]{Sembach+03,Fox+04} indicating the presence of a more highly ionized and hotter medium at $10^{5-6}$ K, likely due to turbulent mixing at the interface between the HVCs and the corona \citep[e.g.][]{Kwak+11}. In recent years, lines of different ions (e.g. \cii, \ciii, \siii, \siiii), probing low ionization state material at $T \sim 10^{4-5}$ K, have been observed in absorption towards extragalactic sources in the Milky Way's halo \citep[e.g.][]{Shull+09,Lehner&Howk+11}. They fill about $70-90 \%$ of the sky and are often associated with the neutral gas emission from HVCs \citep[e.g.][]{Lehner+12}.

Beyond the Milky Way, \hi\ emission from the CGM is harder to detect due of its low column densities. However, nearby galaxies observed with enough sensitivity have systematically revealed the presence of gas complexes and layers of so-called extraplanar gas that typically extend up to a few~$\sim10$ kpc above and below the plane of the galaxy discs \citep[e.g.][]{Oosterloo+07, Gentile+13}.
Extraplanar gas layers likely consist of individual clouds whose origin may be internal or external to the galaxy \citep{Fraternali02, Boomsma+08}.
Their masses, sizes and location closely resemble those of the largest HVCs of the Milky Way \citep[see discussion in][]{Sancisi+08}. In the Local Group, M\,31 has a population of \hi\ clouds with masses down to $\sim 10^5$ $\mo$ and located at distance of tens of kpc from the galaxy \citep{Thilker+04, Lewis+13}. 

Recently, the COS-Halos survey has constrained the physical properties of the CGM in nearby galaxies \citep[e.g.][]{Werk+12, Tumlinson+13}, detecting gas through absorption lines against background QSO spectra for a sample of galaxies in the low-redshift ($0.1<z<0.35$) Universe. The picture that has emerged is that a large fraction of the sample galaxies, regardless of their type, are surrounded by a cool and ionized circumgalactic phase (T $ \lesssim 10^5$ K): both $\Lya$ absorbers and low/intermediate ionization elements (\cii, \ciii, \mgii, \siii, \siiii) have shown strong column densities out to projected distances of 150 kpc from the galaxy centre \citep{Tumlinson+13, Werk+13}. Detections of highly ionized material (e.g. \ovi) have also been reported \citep{Tumlinson+11, Werk+14}. The amount of \ovi\ around star-forming galaxies is larger than around galaxies with little or no star formation \citep{Tumlinson+11}, indicating that the warmer CGM reflects the bimodality of the two types of galaxies.

The aim of this paper is to shed light on the coexistence and the ubiquity of different gas phases in the CGM, and, in particular, to understand under what conditions the cool gas can survive in the hot corona. Our working hypothesis is that the observed cool/warm and ionized phase is associated with the interaction and mixing between the hot coronal medium and a cooler neutral phase. The mechanism originating this cool phase can be various: for instance the Milky-Way's HVCs, gas streams due to tidal/ram pressure stripping (e.g. the MS) or, potentially, clouds ejected by powerful outflows \citep[e.g. the HVC toward the Large Magellanic Cloud,][]{Barger+16}. In this paper, we do not investigate the origin of the cool clouds but focus on the problem of their survival. In Sec. \ref{Hydrodynamical simulations} we introduce the set of hydrodynamical simulations performed justifying the choices of the parameters and we briefly describe the main features of the code that we used. In Sec. \ref{Results} we present the results of our simulations and we compare them with the COS-Halos observations. In Sec. \ref{Discussion} we discuss our simulations by referring to other numerical works investigating the survival of cool clouds, while in Sec. \ref{Conclusions} we summarize our main results. 

\section{Hydrodynamical simulations}
\label{Hydrodynamical simulations} 
We performed a suite of two-dimensional hydrodynamical simulations of cool neutral clouds (T $=10^4$ K) travelling through a hot and low-density coronal medium at given initial velocities (see Sec.~\ref{Discussion1} for a discussion about the possible limitations of the 2D hydrodynamical simulations). Hydrodynamical instabilities (both Kelvin-Helmholtz and Rayleigh-Taylor instabilities) tear up the cool cloud during its motion, producing a long turbulent wake where cloud and corona materials mix efficiently producing gas at intermediate temperature \citep[see][]{Murray+93,Heitsch&Putman09,Kwak+11}. This gas mixture may evaporate in the surrounding hot medium or, if radiative cooling is effective, condensate in cooler and smaller structures \citep{Marinacci+10,Armillotta+16}.
Hereafter we distinguish three ranges of temperature:
\begin{itemize}
\item \textit{cool gas}: gas at $T<10^5$ K, both neutral and ionized, responsible for the absorptions in \hi\ and low-intermediate ionization-state elements; 
\item \textit{warm gas}: gas at $10^5<T<10^6$ K, responsible for the absorptions in \ovi\ and other high ionization-state elements;
\item \textit{hot gas}: gas at $T>10^6$ K, typical of a galactic corona at the virial temperature.
\end{itemize}
The main goal of our work is to understand, by sampling different cloud sizes and velocity, whether the cool gas may survive its journey through the coronal medium. When we refer to the cool gas, we take into account both the cool gas belonging to the remnant of the initial cloud, and the cool gas present in the turbulent wake.

\begin{table}
\centering
\begin{tabular}{cccccc}
\hline 
\hline
$T_\mathrm{cor}$& $n_\mathrm{cor}$& $Z_\mathrm{cor}$& $T_\mathrm{cl}$& $n_\mathrm{cl}$& $Z_\mathrm{cl}$\\
(K) & (cm$^{-3}$) & (Z$_\mathrm{\odot}$) &(K) & (cm$^{-3}$) &  (Z$_\mathrm{\odot}$)\\
\hline
$2\times10^6$& $10^{-4}$&$0.1$ &$10^4$&$2\times10^{-2}$&$0.3$\\
\hline
\hline
\end{tabular} 
\caption{Initial parameters of all our simulations: coronal temperature $T_\mathrm{cor}$, coronal density $n_\mathrm{cor}$, coronal metallicity $Z_\mathrm{cor}$, cloud temperature $T_\mathrm{cl}$, cloud density $n_\mathrm{cl}$, cloud metallicity $Z_\mathrm{cl}$.}
\label{FixedParameters}
\end{table}

The parameters fixed in each simulation are listed in Table \ref{FixedParameters}. In all simulations the parameters of the coronal medium were chosen assuming the typical properties of the Milky Way at large distances from the galactic disc (50-150 kpc, following the observed impact parameters of the COS-Halos survey). The coronal temperature was set to $2\times10^6$ K, roughly the coronal temperature of the Milky Way \citep{Fukugita&Peebles06, Miller&Bregman15}, while the coronal metallicity was set to $0.1$ Z$_\mathrm{\odot}$, according to the value estimated for those galaxies in which the hot halo was actually observed in X-rays \citep{Bogdan+13, Hodges-Kluck&Bregman13, Anderson+16}. For the Milky Way the value is not well constrained, but studies through both Far Ultraviolet absorption spectra and emission lines of \ovii\ and \oviii\ returned values between $0.1$ and $0.3$ Z$_\mathrm{\odot}$ \citep{Sembach+03, Miller&Bregman15}. We assumed a coronal number density of $10^{-4}$ cm$^{-3}$, this value is compatible with the value of $\sim 10^{-4}$ cm$^{-3}$  necessary to explain the multiphase medium at the MS location \citep[e.g.][]{Fox+05, Kalberla+06}, the average value of $\sim 2\times10^{-4}$ cm$^{-3}$ within $100$ kpc found by \citet{Grcevich&Putman09},  the coronal number density of $1.3-3.6\times10^{-4}$ cm$^{-3}$ in the range $50-90$ kpc from the Galactic disc found by \citet{Gatto+13} and the upper limit of $\sim 10^{-4}$ cm$^{-3}$  found by \citet{Besla+12} by reconciling simulations of MS formation and observations. 

We assumed pressure equilibrium between the cloud and the external medium. Our numerical experiments have indeed shown that if the cloud is initially out of pressure equilibrium with the ambient medium it readjusts itself and it reaches pressure equilibrium in a few Myr. With this prescription the cloud number density is fixed by environmental parameters to $2\times10^{-2}$ cm$^{-3}$. The cloud metallicity was fixed to 0.3 Z$_\mathrm{\odot}$ in all simulations. This value is in agreement with the metallicities measured for most high-velocity complexes in the Milky Way, typically between 0.1 and 0.5 Z$_\mathrm{\odot}$ \citep[e.g.][]{Wakker01, Collins+07, Shull+11}. We point out that a value of 0.3 Z$_\mathrm{\odot}$ may indicate that the metallicity is not primordial and it could have been enhanced from the star-forming disc \citep{Fraternali+15, Marasco+17}. A metallicity of 0.3 is also representative of the Magellanic Clouds \citep{Hunter+09}.

\begin{table}
\centering
\begin{tabular}{cccc}
\hline 
\hline
Sim.&$v_\mathrm{cl}$& $r_\mathrm{cl}$& $M_\mathrm{cl}$\\ 
&($\kms$) &  (pc)& ($\mo$)\\
\hline
1&$100$&$50$&$1.6\times10^2$\\
2&$200$&$50$&$1.6\times10^2$\\
3&$300$&$50$&$1.6\times10^2$\\
4&$100$&$100$&$1.3\times10^3$\\
5&$200$&$100$&$1.3\times10^3$\\
6&$300$&$100$&$1.3\times10^3$\\
7&$100$&$250$&$2\times10^4$\\
8&$200$&$250$&$2\times10^4$\\
9&$300$&$250$&$2\times10^4$\\
10&$100$&$500$&$1.6\times10^5$\\
11&$200$&$500$&$1.6\times10^5$\\
12&$300$&$500$&$1.6\times10^5$\\
\hline
\hline
\end{tabular} 
\caption{List of the performed simulations. We varied both the initial cloud velocity, $v_\mathrm{cl}$, and the initial cloud radius, $r_\mathrm{cl}$. Since the cloud number density is fixed in all the performed simulations, different initial cloud radii correspond to different initial cloud masses, $M_\mathrm{cl}$.}
\label{ChangedParameters}
\end{table}

The values of the parameters that characterize the different simulations are listed in table \ref{ChangedParameters}. The two parameters really changing in our set of simulations are the initial cloud velocity with respect to the hot gas, $v_\mathrm{cl}$, and the initial cloud radius, $r_\mathrm{cl}$. Our goal is to investigate how these two physical quantities influence the cloud survival. We assumed three different values for the initial velocity, 100, 200 and 300 $\kms$. These values are compatible with the results found by \citet{Tumlinson+13} and \citet{Werk+13}: the material detected by the COS-Halos observations is within approximately $\pm 200 \kms$ of the galaxy systemic velocity. This range also encompasses the typical rotational velocities for not-extremely elongated orbits in Milky Way-like halos \citep[e.g.,][]{Lux+13}. The values chosen for the initial radius of the cloud are 50, 100, 250 and 500 pc, corresponding to cloud masses of $1.6\times10^2$, $1.3\times10^3$, $2\times10^4$ and $1.6\times10^5$ M$_\mathrm{\odot}$, respectively. The estimated masses of most of the Milky Way HVCs, for which good distance constraints exist, lie in a range of higher masses, between $10^5$ and $5\times10^6 \mo$ \citep{Putman+12}. However, these complexes are always composed by smaller clouds \citep{Thom+08,Hsu+11}, then the masses which we chose for our simulations are fully justified. Also the MS shows clear evidence for a hierarchy of structures, in the form of cloudlets around the main filaments of the Stream \citep{Putman+03,Fox+14}. The same cloud sizes are observed around external galaxies, in particular M\,31 \citep{Thilker+04}. 

In our simulations the clouds are initially spherical. This assumption is unrealistic because the geometry of the \hi\ clouds around our Galaxy is strongly irregular \citep[e.g.][]{Putman+12}. In order to make our simulations more realistic, we allowed for a quick breakdown of the sphericity and homogeneity of the cloud by introducing internal random motions inside it. Both in most Milky Way HVCs and in the MS, the typical velocity dispersion, $\sigma$, observed for the gas phase at $T \lesssim 10^4$ K is $\sim 12 \kms$ \citep[e.g.][]{Kalberla+06}. This dispersion is mainly due to turbulent motions, rather than to thermal broadening. We introduced turbulence inside the cloud both along the x-axis and the y-axis: the initial velocity inside the cloud follows a gaussian distribution with dispersion $10 \kms$ for both the axes and mean value $0 \kms$ for the y-axis and the initial cloud velocity, $v_\mathrm{cl}$, for the x-axis, which is the cloud direction of motion \citep[see also][]{Armillotta+16}.

\subsection{The numerical scheme}
\label{The numerical scheme}

All simulations were performed with the ATHENA code \citep{Stone+08}, using a two-dimensional Cartesian geometry with fixed grid. The 2D geometry suppresses one of the dimensions perpendicular to the cloud velocity and it results in simulating an infinite cylinder that is moving perpendicular to its long axis. From the simulations we obtained quantities per unit length of the cylinder, then, to relate these to the corresponding quantities for an initially spherical cloud of radius $r_\mathrm{cl}$, we multiplied the cylindrical results by the length $4r_\mathrm{cl}/3$ within which the mass of the cylinder equals the mass of the spherical cloud. We used this correction to calculate the gas mass below a given temperature for all the simulations presented in this work \citep[see also][]{Marinacci+10}.

We implemented an algorithm to make the grid adaptive and moving, in order to follow the cloud along its motion. This is necessary in order to reduce the size of the computational domain and the time needed to run the simulations. After every Myr the algorithm repositions the cloud's head at the grid centre. Open boundary conditions were imposed at the four sides of the simulation domain. When the cloud head is repositioned at the grid centre, new gas is added in the grid portion towards which the cloud is moving. We force this gas to assume the same physical conditions of the unperturbed hot corona.
The spatial resolution is 2pc x 2pc. Our results on the study of the code convergence showed that this resolution appear to be optimal for limiting divergence problems related to numerical diffusion \citep[see][for details]{Armillotta+16}. 

We did not take into account gravitational acceleration and coronal density variation along the cloud motion. Including a gravitational field is indeed beyond our current purposes. We were interested to study the average environmental effects on the cloud survival so we did not specify its trajectory (outflow or inflow) or its exact location in the circumgalactic halo. Furthermore, we neglected the presence of self-gravity inside the cloud. The effect of self-gravity is to stabilize the cloud against the formation of hydrodynamical instabilities, slowing down its ablation \citep{Murray+93}. However, in our simulations, the virial ratio of the clouds ranges from a few $\times~10^4$ (clouds with $r_\mathrm{cl}= 50$ pc) to a few $\times~10^2$ (clouds with $r_\mathrm{cl}= 500$ pc). In all cases, it is orders of magnitude larger than the unity, then we expect that self-gravity plays a minor role in our simulations.

We included thermal physics in our simulations by adding terms to take into account radiative cooling/heating and thermal conduction to the energy equation:
\begin{equation}
\dfrac{\partial e}{\partial t} + \mathbf {\nabla} \cdot [ (e+P) \mathbf{v}]= \,-\, \rho^2 \Lambda_\mathrm{net} - \mathbf {\nabla} \cdot \textbf{\textit{q}}
\label{Energy}
\end{equation}
where \textit{e} is the energy density, \textbf{v} the velocity, \textit{$\rho$} the density and $P=(\gamma-1) U$ the pressure with $U$ the internal energy density and $\gamma=5/3$, \textbf{\textit{q}} is the so-called `heat conduction flux', $\Lambda_\mathrm{net}$ is the net cooling/heating rate ($\Lambda_\mathrm{net}= \Lambda - \Gamma$, where $\Lambda$ and $\Gamma$ are, respectively,  the  radiative  cooling and the photo-heating rate).
Despite modules for these processes were present in ATHENA, we modified them to make the code more suitable for our purposes, as explained below. 

\subsubsection{Radiative processes}

The effects of radiative cooling and heating were included in the code by using the cooling and heating rates calculated through the CLOUDY spectral synthesis code \citep[version c13; ][]{Ferland+13} and tabulated across a range of physical conditions. We evaluated the effect of photoionization from a uniform UV background at $z=0.2$ (the average redshift of the COS-Halos sample of galaxies), accounting for the UV radiation emitted by all stars and AGN throughout the evolution of the Universe and attenuated by the $\Lya$ forest \citep{Haardt&Madau12}. Collisional ionization of all atoms and ions is also included in the model. All chemical elements were assumed to be in ionization equilibrium, which means that atomic processes (recombination, photoionization, collisional ionization) become steady in a timescale lower than the hydrodynamical timescale. 

Cooling and heating rates were tabulated as a function of total gas temperature T, metallicity Z, and hydrogen numerical density n$_\mathrm{H}$. T ranges from $10^{4}$ to $10^{7}$ K with a resolution of 0.05 dex in the log space. n$_\mathrm{H}$ ranges from $10^{-6}$ to $1$ cm$^{-3}$ with a resolution of 0.5 dex. The gas metallicity, Z, assumes only two values, $0.3$ Z$_\mathrm{\odot}$ and $0.1$ Z$_\mathrm{\odot}$. The abundances at ${Z =}$ Z$_\mathrm{\odot}$ and ${Z = 0.1}$ Z$_\mathrm{\odot}$ were taken from \citet{Sutherland&Dopita93}. For the $Z=0.3$ Z$_\mathrm{\odot}$ model, we  linearly interpolated the abundances at ${Z =}$ Z$_\mathrm{\odot}$ and ${Z = 0.1}$ Z$_\mathrm{\odot}$ on logarithmic scale. 

\begin{figure}
\includegraphics[width=0.48\textwidth]{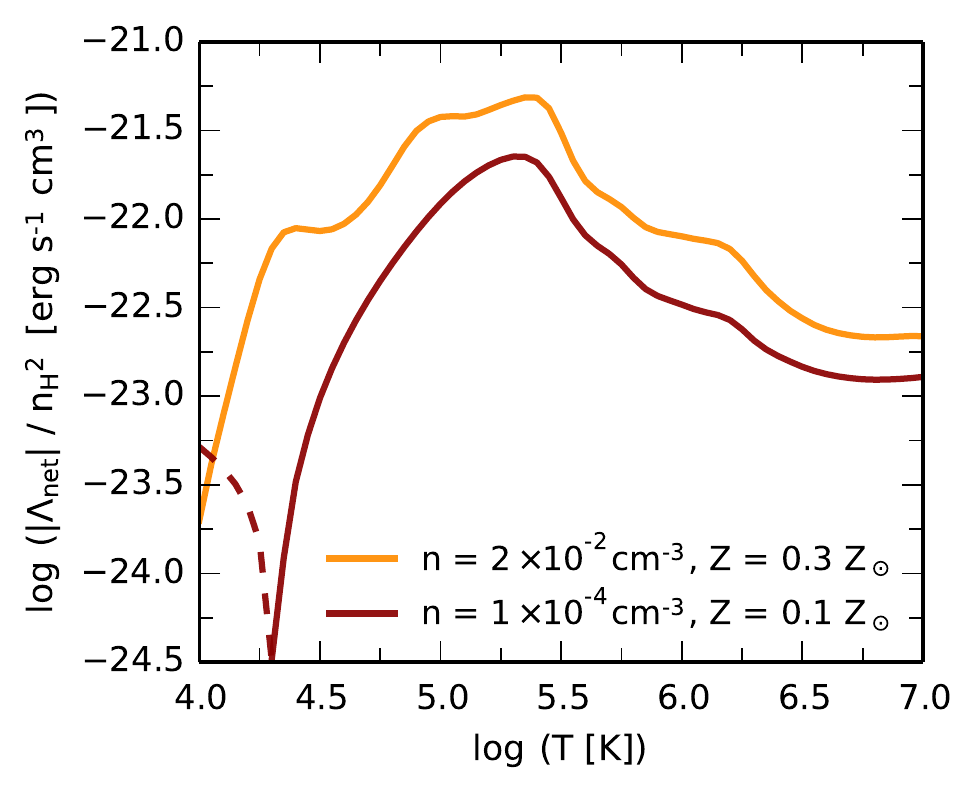}
\caption{Absolute value of net cooling/heating rate as a function of gas temperature for $n=2\times10^{-2}$ cm$^{-3}$ and $Z=0.3$ $Z_\mathrm{\odot}$ (orange line) and $n=10^{-4}$ cm$^{-3}$ and $Z=0.1$ Z$_\mathrm{\odot}$ (brown line). We plot the absolute value of the radiative cooling minus the photoheating rate per unit volume, divided by $n_\mathrm{H}^2$. We use a dashed line to indicate that at lower gas density and metallicity heating dominates over cooling for $T< 10^{4.3}$ K and the net effect is heating.}
\label{CLOUDY}
\end{figure}

Since the tabulated cooling/heating rates depend on n$_\mathrm{H}$, we need to know the value of n$_\mathrm{H}$ in each cell of the computational domain and at each time step. For this reason, we calculated the temporal and spatial evolution of n$_\mathrm{H}$ by modelling it as a passive scalar field. Numerically, we treated advection of hydrogen mass in the same manner as advection of mass. The initial conditions were calculated in each cell by multiplying the total gas density by the hydrogen abundance at given metallicity. 

The results of the photoionization model are presented in Fig. \ref{CLOUDY}. The plot shows the net cooling/heating rates as a function of temperature in the log scale for two extreme couple of values present in our simulations, one with density and metallicity of the cloud ($n=2\times10^{-2}$ cm$^{-3}$ and $Z=0.3$ $Z_\mathrm{\odot}$) and the other with density and metallicity of the coronal medium ($n=10^{-4}$ cm$^{-3}$ and $Z=0.1$ Z$_\mathrm{\odot}$). Most of our cells falls typically within these values. At $T\sim 10^{4.3}$ K the curve at low density and metallicity show a discontinuity, below which heating dominates over cooling. We plotted this region using a dotted line, in order to distinguish it from the region where cooling dominates.

\subsubsection{Thermal conduction}

According to the classical theory, the heat conduction flux is given by the Spitzer formula \citep{Spitzer62}:
\begin{equation}
\textbf{\textit{q}} = - \kappa_\mathrm{Sp} \cdot \mathbf {\nabla} T \:,
\label{Spitzer}
\end{equation}
where ${\nabla} T$ is the temperature gradient, and the heat conduction coefficient is
\begin{equation}
\kappa_\mathrm{Sp} = \dfrac{1.84 \times 10^{-5} }{\mathrm{ln\Psi}}  \cdot T^{5/2}\: \: \: \mathrm{erg \,s^{-1} \,K^{-1}\, cm^{-1}} \:,
\label{Spitzer2}
\end{equation}
where $\ln\Psi$ is the Coulomb logarithm and it can be expressed as 
\begin{equation}
\mathrm{ln\Psi} = 29.7 + \mathrm{ln} \left[ \dfrac{T_{e} / 10^6 K}{\sqrt{n_{e} / cm ^{-3}}}\right]
\end{equation}
with $n_\mathrm{e}$ being the electron density and $T_\mathrm{e}$ the electron temperature.

The classical prescription breaks down in the presence of a magnetic field and the efficiency of thermal conduction may be strongly reduced \citep[e.g.][]{Chandran&Cowley98}. To take into account this effect we assumed an efficiency of 10\% for the Spitzer thermal conduction (Eq. \ref{Spitzer}). This value is in agreement with the result found by \citet{Narayan&Medvedev01}, which studied the thermal conduction suppression when the magnetic turbulence extends on a wide range of length scales, as it might happen with strong-intermediate MHD turbulence. Furthermore, the efficiency of the classical thermal conduction can be reduced when the local temperature scale-length falls below the mean free path of the conducting electrons. In this case the heat flux is replaced by a flux-limited form the so-called `saturated heat flux' \citep{Cowie&McKee77}:
\begin{equation}
\vert \,\textbf{\textit{q}}_\mathrm{sat}\vert = 5 \Phi_{\mathrm{s}}\rho c^3\;,
\label{Saturated1}
\end{equation}
with the sound speed \textit{c} and the density $\rho$. $ \Phi_{\mathrm{s}}$ is an efficiency factor less than or of the order of unity, related to the flux-limited treatment and flux suppression due to magnetic fields.

To take into account all these mentioned effects, we modified the Spitzer formula (Eq. \ref{Spitzer}):
\begin{equation}
\textbf{\textit{q}} = - 0.1\,\dfrac{\kappa_\mathrm{Sp} }{1+\sigma} \cdot \mathbf {\nabla} T \;,
\label{Eqcode}
\end{equation}
where $\sigma$ is the absolute ratio between the classical heat flux and the saturated heat flux, according to \citet{Balbus86}. We solved this equation by including a module for isotropic thermal conduction in ATHENA. This module was implemented through an implicit scheme in order to avoid very restrictive CFL constraints on the time step. Further details on the hydrodynamical treatment of thermal conduction and the explanation of the algorithm have been described in \citet{Armillotta+16}.

\section{Results}
\label{Results}

\begin{figure}
\includegraphics[width=0.48\textwidth]{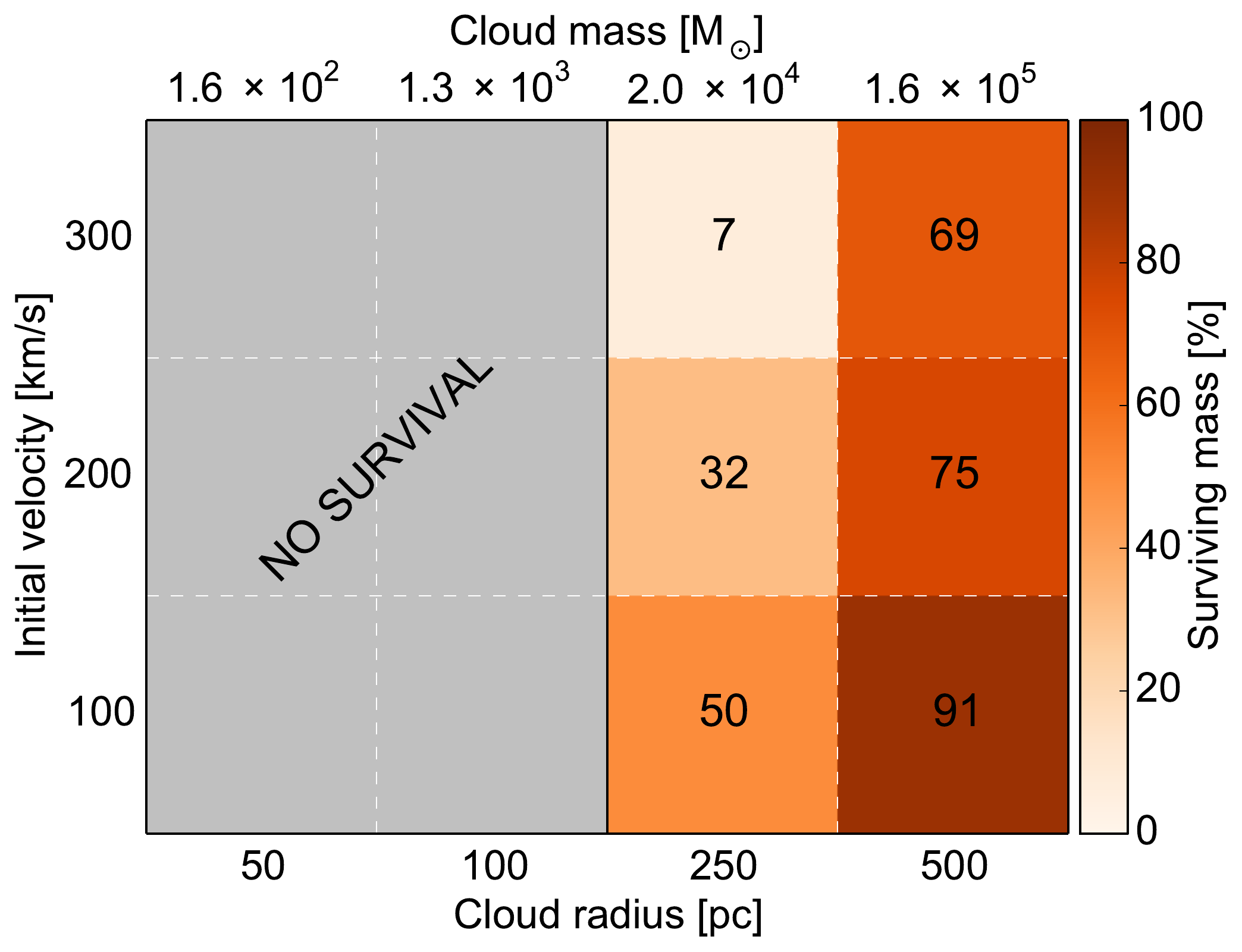}
\caption{Parameter space (radius-velocity-mass) explored in our simulations to test cool gas survival for 250 Myr. The number written in each square and the color bar indicate the percentage of cool gas (T $\leq 10^5$ K) that survives. The grey band represents the range of parameters in which the cloud is not able to survive.}
\label{Fig1}
\end{figure}

We run our simulations of cool clouds interacting with the surrounding coronal gas for 250 Myr.
We found that the mass of cool gas decreases with time in all the performed simulations. As mentioned in Sec.\ref{Hydrodynamical simulations}, the cool gas, stripped from the cloud by hydrodynamical instabilities, mixes with the coronal gas in a turbulent wake. In all the analyzed cases, the wake density is not high enough to allow for gas condensation \citep[like in][]{Armillotta+16}, and most of cool gas lost from the cloud evaporates in the hot coronal medium. However, in several cases, a consistent fraction of cool gas survives until the end of the simulation: most of this gas resides inside the cloud head and only a very small fraction is located in the wake, as we show in Sec \ref{The role of thermal conduction}. 

Fig.~\ref{Fig1} shows the values of initial radius and velocity of the cloud that allow the cool gas survival for at least 250 Myr, and the fraction of gas able to survive after this time. Clouds with radius equal to or less than 100 pc are entirely destroyed after 250 Myr. Clouds with radius equal to 250 pc are able to survive. However, the fraction of surviving mass strongly depends on the initial cloud velocity, decreasing with increasing velocity. The cloud with $v_\mathrm{cl}=100 \kms$ is able to keep $\sim50\%$ of its own initial mass after 250 Myr, while the cloud with $v_\mathrm{cl}=300 \kms$ retains only $\sim 7\%$. The situation changes completely for larger clouds ($r_\mathrm{cl}=500$ pc): a significant fraction of cool gas survives ($\gtrsim 70 \%$) and its dependence on the initial velocity becomes rather weak. 

\begin{figure}
\includegraphics[width=0.48\textwidth]{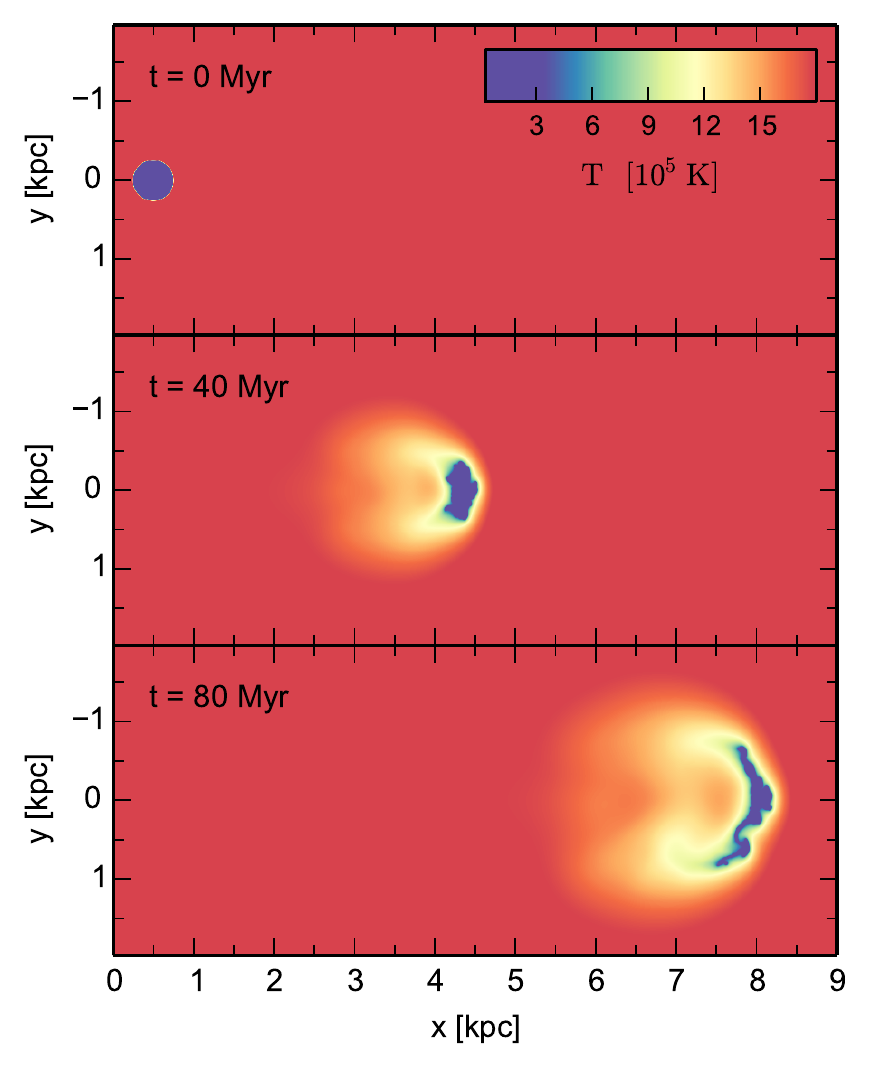}
\caption{Temperature snapshots of the simulation with initial cloud velocity $100\kms$ and initial cloud radius $250$ pc (Sim.~7 in Tab.~\ref{ChangedParameters}) with thermal conduction. The time at which the snapshots have been taken is indicated in each panel.}
\label{Fig1.2}
\end{figure}

\begin{figure*}
\includegraphics[width=0.9\textwidth]{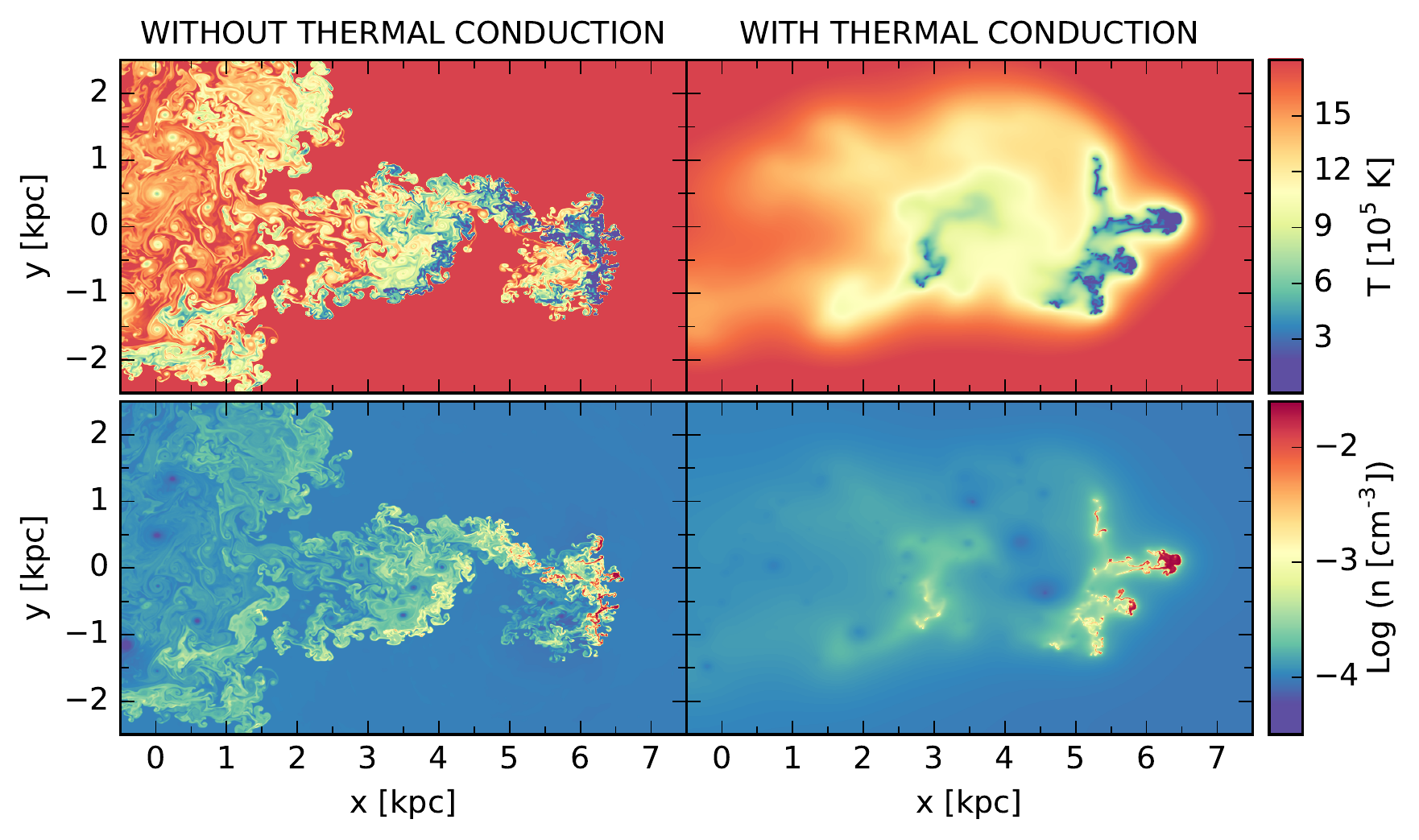}
\caption{Temperature (top panels) and number density (bottom panels) snapshots of the simulation with initial cloud velocity $100\kms$ and initial cloud radius $250$ pc (Sim.~7 in Tab.~\ref{ChangedParameters}) without (left panels) and with (right panels) thermal conduction. The time at which the snapshots have been taken is 200 Myr.}
\label{Fig2}
\end{figure*}
\begin{figure}
\includegraphics[width=0.48\textwidth]{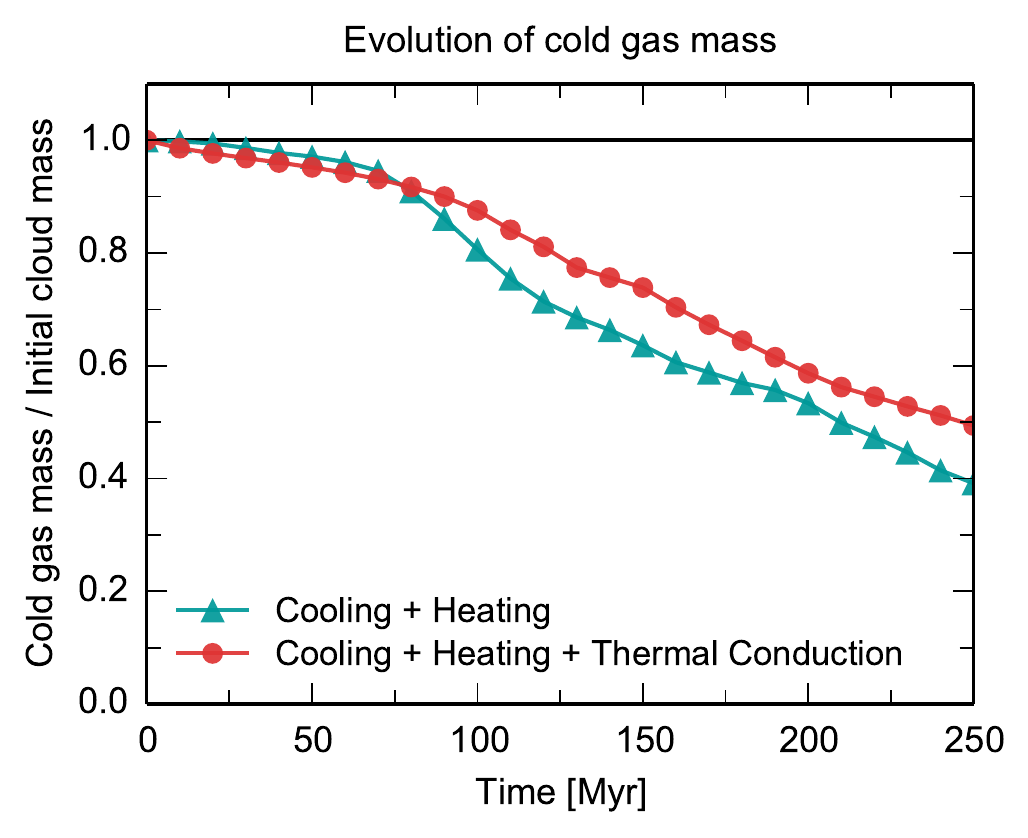}
\caption{Evolution of the mass of cool gas ($T<10^5$ K) with time for two simulations with initial cloud velocity $100\kms$ and initial cloud radius $250$ pc (Sim.~7 in Tab.~\ref{ChangedParameters}): one without (green line) and one with thermal conduction (red line).}
\label{Fig3}
\end{figure}

A key issue in this analysis is the efficiency of coronal ram pressure to warp the cloud, triggering its ablation. As an example, in Fig. \ref{Fig1.2} we show the temperature distribution on the grid at the initial time (top panel) and after 40 and 80 Myr (middle and bottom panel respectively) for the simulations with initial cloud velocity $100\kms$ and initial cloud radius $250$ pc (Sim.~7 in Tab.~\ref{ChangedParameters}). The coronal flow flattens the cloud perpendicularly to the motion direction, increasing its cross section as a consequence of the Bernoulli effect. The cloud distortion triggers the formation of Kelvin-Helmholtz instabilities resulting from the shear flow over its leading face and Rayleigh-Taylor instabilities along its symmetry axis due to deceleration exerted from the coronal pressure. The development of these two instabilities lead to the cloud ablation. Therefore, the time scale associated to the cloud survival is strongly related to time scale associated to ram pressure exerted on the cloud by the coronal medium, the so-called drag time \citep[e.g.][]{Fraternali&Binney06}:
\begin{equation}
t_\mathrm{drag}=\dfrac{M_\mathrm{cl}}{v_\mathrm{cl}\sigma\rho_\mathrm{cor}}\; ,
\end{equation}
where $\sigma \simeq \pi {r_\mathrm{cl}}^2$ is the cross section of the cloud. The drag time increases with increasing cloud size ($t_\mathrm{drag}\propto{M_\mathrm{cl}}/{\sigma}\propto r_\mathrm{cl}$) and decreases with increasing cloud velocity. The effect of coronal ram pressure on the largest clouds is very weak and they are able to survive for very long times. 

The velocity can also play an important role. Fig. \ref{Fig1} shows that the fraction of surviving mass decreases with increasing velocity for the clouds with $r_\mathrm{cl}=250$ pc and for the clouds with $r_\mathrm{cl}=500$ pc and $v_\mathrm{cl}=100 \kms$ and $200 \kms$.  
A dependence on the velocity is instead roughly absent for the cloud with $r_\mathrm{cl}=500$ pc and $v_\mathrm{cl}\geqslant200 \kms$. The cloud with $v_\mathrm{cl}=300 \kms$ keeps $\sim70\%$ of its own initial mass, while the cloud with $v_\mathrm{cl}=200 \kms$ keeps $\sim 75\%$, a small difference. In this last case the velocity is supersonic (the sound speed is $\sim200\kms$ at $T=2\times10^6$ K) and the radius is large enough to keep the velocities supersonic for almost the entire simulation time. In fact, in our simulations the cloud velocity evolves with time approximately as \citep[][]{Fraternali&Binney06}:
\begin{equation}
v_\mathrm{cl}(t)\simeq \dfrac{v_\mathrm{cl}(0)}{1+t/ t_\mathrm{drag}} \; .
\end{equation}
For $t\ll t_\mathrm{drag}$, $dv_\mathrm{cl}(t)/ dt \propto 1/t_\mathrm{drag} \propto 1/R_\mathrm{cl}$, then, the larger the cloud radius the slower the cloud deceleration. In the supersonic case the formation of a bow shock in front of the cloud suppresses the formation of Kelvin-Helmholtz instabilities and hampers the cloud destruction \citep[see also e.g.][]{Scannapieco&Bruggen15}.

\subsection{The role of thermal conduction}
\label{The role of thermal conduction}

Thermal conduction is a key mechanism to consider in the interaction between different gas phases because it allows for transfer of heat from a hot to a cool medium, accelerating their mixing.

In order to better understand how thermal conduction can influence the cloud survival, we chose a fiducial simulation,  $r_\mathrm{cl}=250$ pc and $v_\mathrm{cl}=100 \kms$ (Sim.~7 in Tab. \ref{ChangedParameters}), and we analyzed it both in the absence and the presence of thermal conduction. Fig.\ref{Fig2} shows the temperature (top panels) and density (bottom panels) distributions on the grid after 200 Myr for the simulations without (left panels) and with (right panels) thermal conduction. The general evolution of the cloud is very different in the two cases. In the absence of thermal conduction the initial cloud is totally destroyed while the turbulent wake is characterized by the presence of a gradient in temperature and in density extended to several kpc downstream in the coronal medium. Cooler and denser regions are situated in the head of the wake, where the gas lost from the cloud is not yet well mixed with the coronal gas. Most of this cool gas is approximately an order of magnitude less dense than the initial cloud ($n\sim10^{-3}$ cm$^{-3}$), then, despite it encompasses a large volume, its contribution to the total mass will be low. Instead, in the presence of thermal conduction the bulk of the initial cloud is partially intact after 200 Myr and the cool gas mainly resides inside the dense cloud, while the wake is nearly lacking cool gas and it is composed by an extremely homogeneous mixture at $T \sim 10^{6}$ K and $n \sim 10^{-4}$ cm$^{-3}$. The effect of thermal conduction is indeed to create a more widespread and warmer wake in which temperature and density gradients are smoothed. 

Fig. \ref{Fig3} shows the quantitative results for these two simulations. We compared the evolution of the mass of cool gas with time and we found that it decreases in both cases. The cool gas lost from the cloud evaporates in the coronal medium because the warm wake, produced by the mixing of cool and hot gas, is unable to cool down. As we saw in Fig. \ref{Fig2}, regardless of the presence of thermal conduction, regions in the wake at temperatures close to the peak of the cooling function ($T\sim10^{5.5}$ K) exhibit very low densities ($n\sim10^{-3}$ cm$^{-3}$) and metallicities ($Z<0.3$ Z$_\mathrm{\odot}$). Under these conditions, the condensation of gas \citep[see e.g.][]{Armillotta+16} is hard to occur. The efficiency of condensation strongly depends on gas density and metallicity (see Fig. \ref{CLOUDY}): decreasing density and metallicity, the cooling rates decrease and the effect of heating photoionization becomes more and more important. Therefore, the net result is the loss of cool gas. However, Fig.\ref{Fig3} shows that the reduction of cool gas mass occurs faster without thermal conduction. After 250 Myr the mass of cool gas is $\sim50\%$ of the initial cloud mass in the presence of thermal conduction compared to less than $40\%$ in the absence of it.

Thermal conduction smooths the velocity and density gradients at the interface between the two fluids, hindering the formation of hydrodynamical instabilities and subsequent mixing and making the cloud core more compact  \citep[see also][]{Marcolini+05, Vieser&Hensler07b}. This phenomenon explains the slow destruction of the cloud in the presence of thermal conduction. Once the gas is stripped from the cloud, thermal conduction changes its role, accelerating the heating of cool gas and its evaporation in the coronal medium. As shown in Fig.\ref{Fig2}, cool gas is nearly absent in the warm and rarefied cloud wake in the presence of thermal conduction. 

\subsection{Column densities: comparison with COS-Halos data}

The key observational properties of the COS-Halos detections is given by their column densities. Here we compare the column densities of our fiducial simulation ($R_\mathrm{cl}=250$ pc, $v_\mathrm{cl}=100 \kms$) at 200 Myr with the COS-Halos observations. We estimated the column density of a given ion, $N_\mathrm{X}$, along vertical lines across the simulation box by summing, for each pixel $i$ along the line of sight, the product between the hydrogen column density, $N_{\mathrm{H},i}$, the abundance of the ion with respect to its own species at the temperature and hydrogen number density of the pixel, $X/A \,(T_i,n_{\mathrm{H},i})$, and the abundance of the species with respect to hydrogen at the pixel metallicity, $A/H \,(Z_i)$, as in the following formula:
\begin{equation}
N_\mathrm{X} \,= \, \sum_{i} N_{\mathrm{H},i} \dfrac{X}{A}(T_i,n_{\mathrm{H},i}) \dfrac{A}{H}(Z_i)
\end{equation}
For this calculation, we excluded the gas at $T>10^6$ K, which is the temperature upper limit for the detected \ovi\ 
\citep{Werk+14}. 
The abundance of the ion with respect to its own species, $X/A$, was obtained through the CLOUDY package, by using the same model including photoionization and collisional ionization that we described in Sec.~\ref{The numerical scheme} for the cooling/heating rates. The abundance of the species with respect to hydrogen, $A/H$, at a given metallicity was obtained by linearly interpolating the abundances at ${Z =}$ Z$_\mathrm{\odot}$ and ${Z = 0.1}$ Z$_\mathrm{\odot}$, taken from \citet{Sutherland&Dopita93}, on logarithmic scale. 

\begin{figure*}
\includegraphics[width=\textwidth]{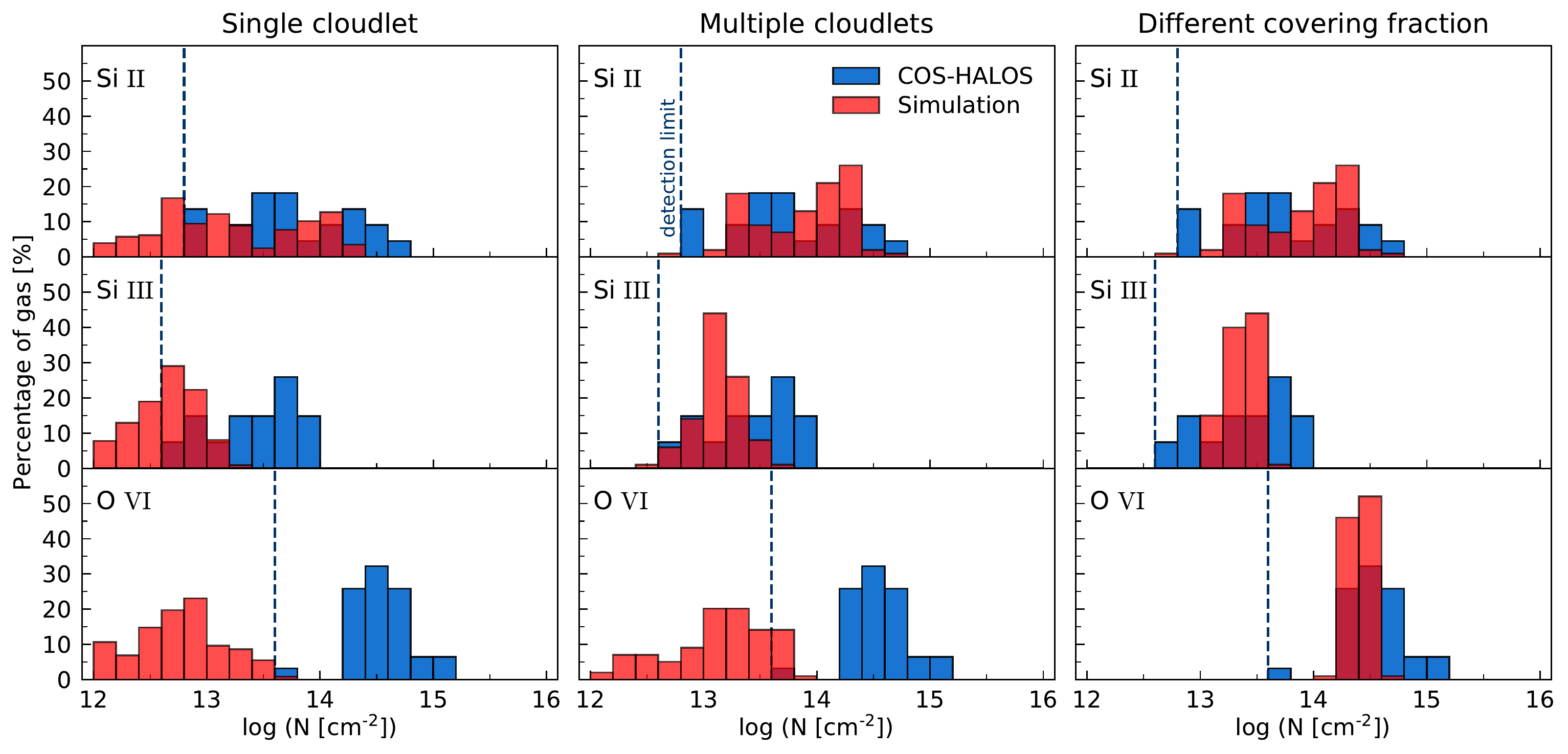}
\caption{Histograms of distribution of the observed (blue bars) and simulated (red bars) data in bins of column density. The top panels show the distribution of \siii, the middle panels the distribution of \siiii\ and the bottom panels the distribution of \ovi. The simulated data were obtained from the fiducial simulation ($R_\mathrm{cl}=250$ pc, $v_\mathrm{cl}=100 \kms$) after 200 Myr. The left panel was obtained by directly comparing the simulation results with the observations. The central panel was obtained by accounting for the contribution of 3 (average detections in COS-Halos) intevening clouds along the line of sight. The right panel was obtained both by accounting for the contribution of 3, 9 and 60 simulated clouds along the line of sight containing respectively \siii, \siiii\ and \ovi. The dotted lines indicate the observational detection limit.}
\label{Fig4}
\end{figure*}

The left panel of Fig.~\ref{Fig4} shows the distribution of \siii, \siiii\ and \ovi\ in bins of column density by directly comparing our simulation predictions to the observations. The blue bars represent the observed data, where we excluded all the upper and lower limits \citep[from][]{Tumlinson+11,Werk+13}. The red bars represent the simulated column densities. The dotted line indicates that the sensitivity limit of the observations, while the simulations allow calculations of much lower values. We found that the ranges of observed and simulated column densities overlay for the low/intermediate ionization elements, while they do not overlap for the \ovi. In the first case the mean values of two distributions differ by a factor of a few, while in the second case they differ by almost two orders of magnitude. Furthermore, the observed column densities extend to a higher range of values. 

\begin{figure}
\includegraphics[width=0.49\textwidth]{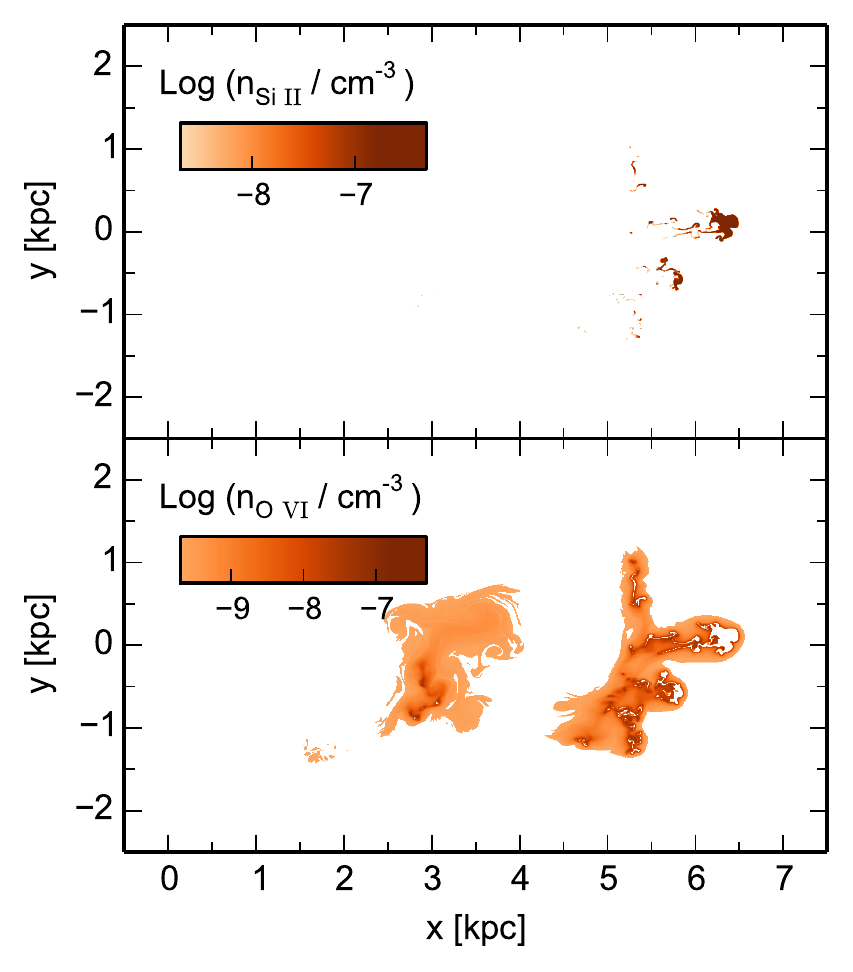}
\caption{Snapshots of \siii\ (top panel) and \ovi\ (bottom panel) number densities for the simulation with initial cloud velocity $100\kms$ and initial cloud radius $250$ pc with thermal conduction. The time at which the snapshots have been taken is 200 Myr. The white color indicates that the ion density is null or negligible (i.e. the cool gas in the cloud head or hot material with $T>10^6$ K, the upper limit of temperature for the detected \ovi).}
\label{Fig5}
\end{figure}

The different column densities may be explained by the fact that the observations can probe multiple clouds along the line of sight, while in our simulations we have effectively followed the evolution of a single cloud. The central panel of Fig. \ref{Fig4}  shows the distribution of \siii, \siiii\ and \ovi\ in bins of column density, accounting for presence of multiple clouds along the line of sight: the new simulated column densities were obtained by summing the column densities along 3 vertical lines taken randomly from the simulation box.
We made this choice because it corresponds to the typical number of kinematic components present in individual absorption lines \citep{Werk+13}. We found that the column densities obtained from the simulation are in very good agreement with the observations for the \siii. However,  the agreement worsens with increasing ionization state. Indeed, the ranges of observed and simulated column densities barely overlap for the \ovi, they differ by about one order of magnitude. 

\begin{figure}
\includegraphics[angle = 270, width=0.52\textwidth]{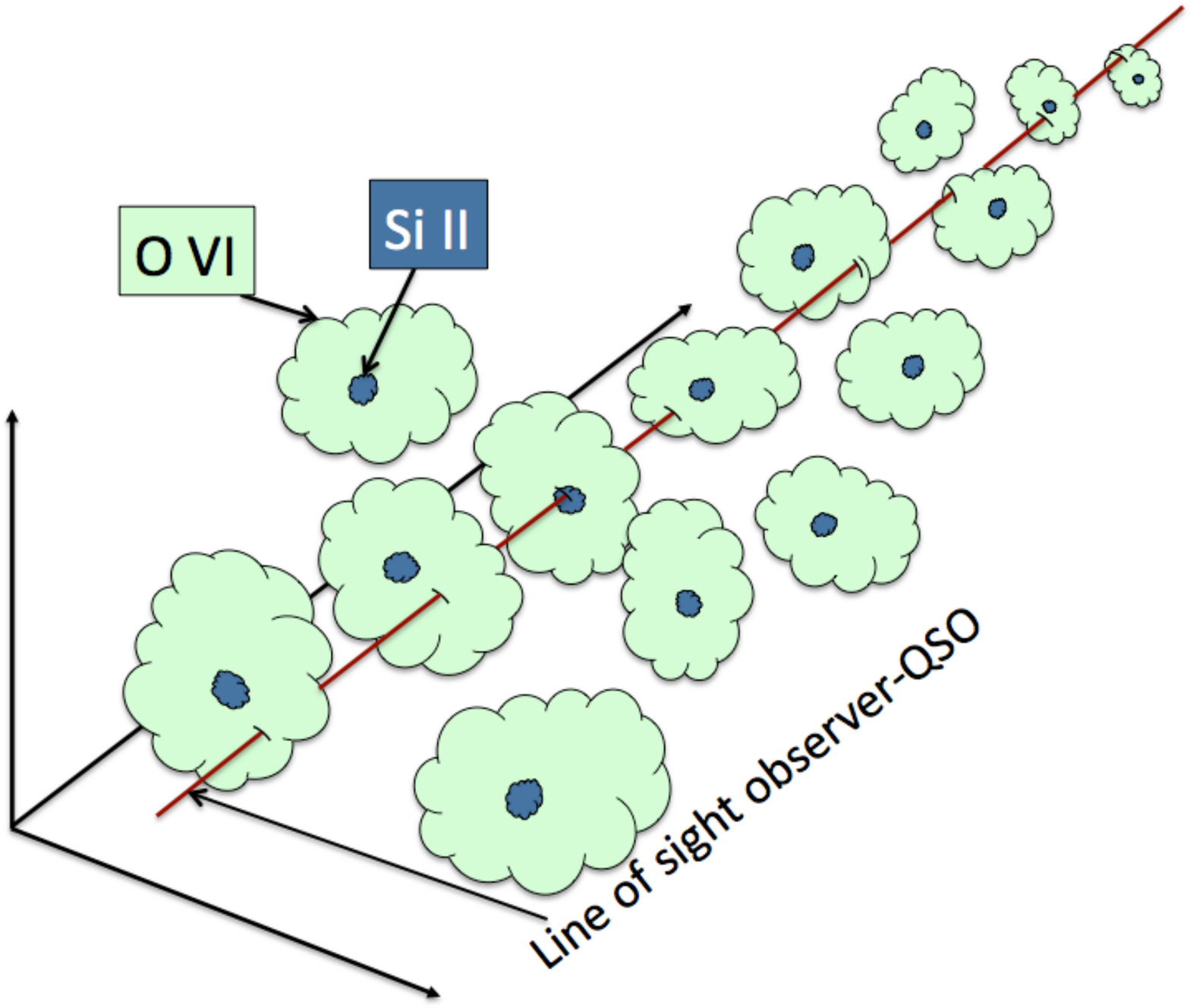}
\caption{Schematic picture of \siii\ and \ovi\ distribution in the CGM. Within each cloud, \siii\ occupies a volume much smaller than \ovi. As a consequence, a generic line of sight (red line) crossing multiple clouds intersects a larger number of \ovi\ absorbers rather than \siii\ absorbers.}
\label{Cartoon}
\end{figure}

\begin{figure}
\includegraphics[width=0.45\textwidth]{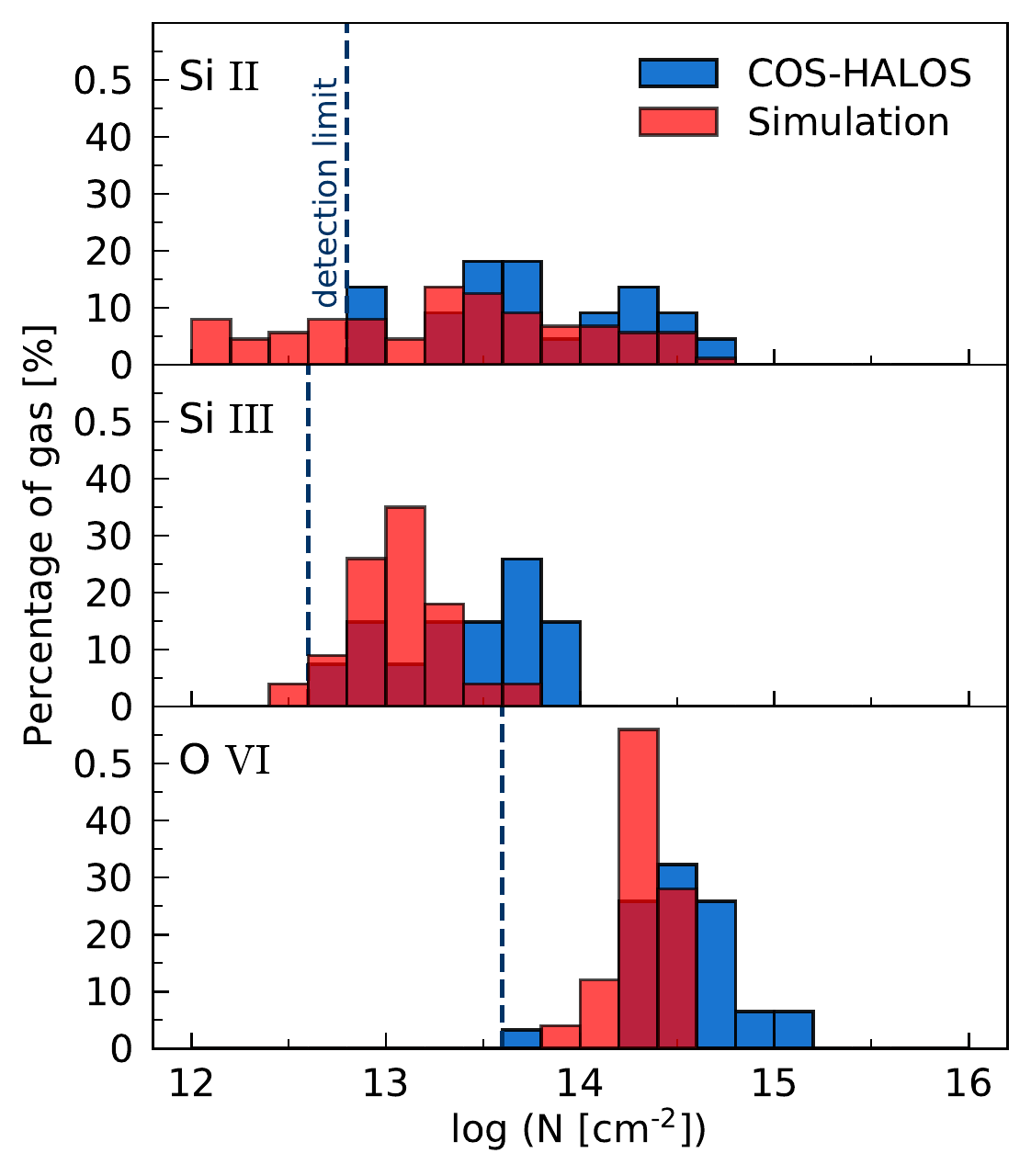}
\caption{Histograms of distribution of the observed (blue bars) and simulated (red bars) data in bins of column density. The top panels show the distribution of \siii, the middle panels the distribution of \siiii\ and the bottom panels the distribution of \ovi.  The simulated data were obtained from the simulation  with $R_\mathrm{cl}=500$ pc, $v_\mathrm{cl}=200 \kms$ (Sim.~11 in Tab.~\ref{ChangedParameters}) after 200 Myr. The histogram of the \siii\ was obtained by directly comparing the simulation results with the observations. The histograms of the \siiii\ and \ovi\ were obtained by considering the presence respectively of 3 and 20 clouds along the line of sight.
The dotted lines indicate the observational detection limit.}
\label{Fig4.2}
\end{figure}

In order to examine the different behaviour of a weakly ionized element, as the \siii, and a highly ionized element, as the \ovi, we analyzed their spatial distribution in our simulations. Fig. \ref{Fig5} shows the \siii\ (top panel) and \ovi\ (bottom panel) number density distributions on the grid after 200 Myr in our fiducial simulation. The comparison between these snapshots and the right panels of Fig. \ref{Fig2} shows that the \siii\ traces the coolest and densest gas phase, it is present in the head of the cloud and in a few cool filaments in the wake, while it is totally absent everywhere else. Instead, the \ovi\ is nearly absent in the coolest regions, while it is widely distributed in the wake, with peaks of density at the boundaries of the cool structures, where the mixing with hotter gas occurs and thermal conduction is stronger. Therefore, the \ovi\ traces a warmer (T $\sim 10^{5-6}$ K) and more widespread gas phase, which occupies regions more spatially extended with respect to cooler gas (see Fig. \ref{Fig5}).
Since our simulations are two-dimensional, we can assume that the 2D cloud in Fig.~\ref{Fig5} is the projection on the sky of a 3D cloud. It is therefore natural to expect that, generic lines of sight should cross an higher number of warm wakes containing \ovi\ rather than cool cloud cores containing low ionization elements. In our simulation we estimated that the ratio between the area occupied by \ovi\ and the area occupied by a generic low ionization element, as the \siii, is $\sim 20$. Therefore, detection of $\sim 3$ clouds containing \siii\ could correspond to detection of $\sim 60$ clouds containing \ovi. Fig.~\ref{Cartoon} shows a schematic picture of the expected 3D distribution of \siii\ and \ovi\ in the CGM.

The right panel of Fig.\ref{Fig4} shows the numerical distribution of \siii, \siiii\ and \ovi\ in bins of column density, taking into account a larger probability to detect \ovi\ and \siiii\ along the line of sight with respect to \siii. In this case the \siii\ column densities are the same as those in the central panel. The \ovi\ column densities were obtained by summing the column densities along 60 vertical lines taken randomly from the simulation box, where only 3 of these 60 lines contain \siii. The \siiii\ column densities were obtained in the same way as \ovi\ accounting for the ratio between the area occupied by \siiii\ and the area occupied by \siii\ is $\sim 3$. We point out that for these calculations we made the double assumption that our 2D cloud is both the projection on the sky and a slice along the line of sight of the 3D cloud. The second assumption is necessary to extract a more realistic density profile along the line of sight, since, strictly speaking, this direction is suppressed in the 2D cartesian geometry of our simulations.
The ranges of column densities are in very good agreement with the observations both for the low-intermediate and for the high ionization elements. 

We also calculated the column densities for the simulation with $r_\mathrm{cl}=500$ pc and $v_\mathrm{cl}=200 \kms$ (Sim.~11 in Tab.~\ref{ChangedParameters}), accounting for the presence of a single cloud containing \siii\ along the line of sight and multiple clouds containing \siiii\ and \ovi, as shown in Fig.~\ref{Fig4.2}. The agreement between simulations and observations is excellent for \siii: the larger radius of the cloud implies a greater amount of cool gas and, then, higher column densities. We found a good agreement between the simulations and the observations also for the \siiii\ and \ovi.

We conclude by noting that due to the way thermal conduction operates, in particular suppressing cool cloudlets in the wake, warm ($10^{5-6}$ K) gas can be found at large distances (several kpc) from the location of the cool material (see Fig.~\ref{Fig2} and Fig.~\ref{Fig5}). This may be relevant in explaining the detections of \ovi\ \citep[see][for other possible explanations]{Werk+16} not directly associated with the classical Galactic HVCs, observed in \hi, or with the Magellanic Stream \citep{Sembach+03, Savage+03}.
Indeed, our simulations exhibit the presence of material at intermediate temperature in the wake for times of the order of hundreds of Myr.

\section{Discussion}
\label{Discussion}

\subsection{Comparison of models}

\begin{figure}
\includegraphics[width=0.49\textwidth]{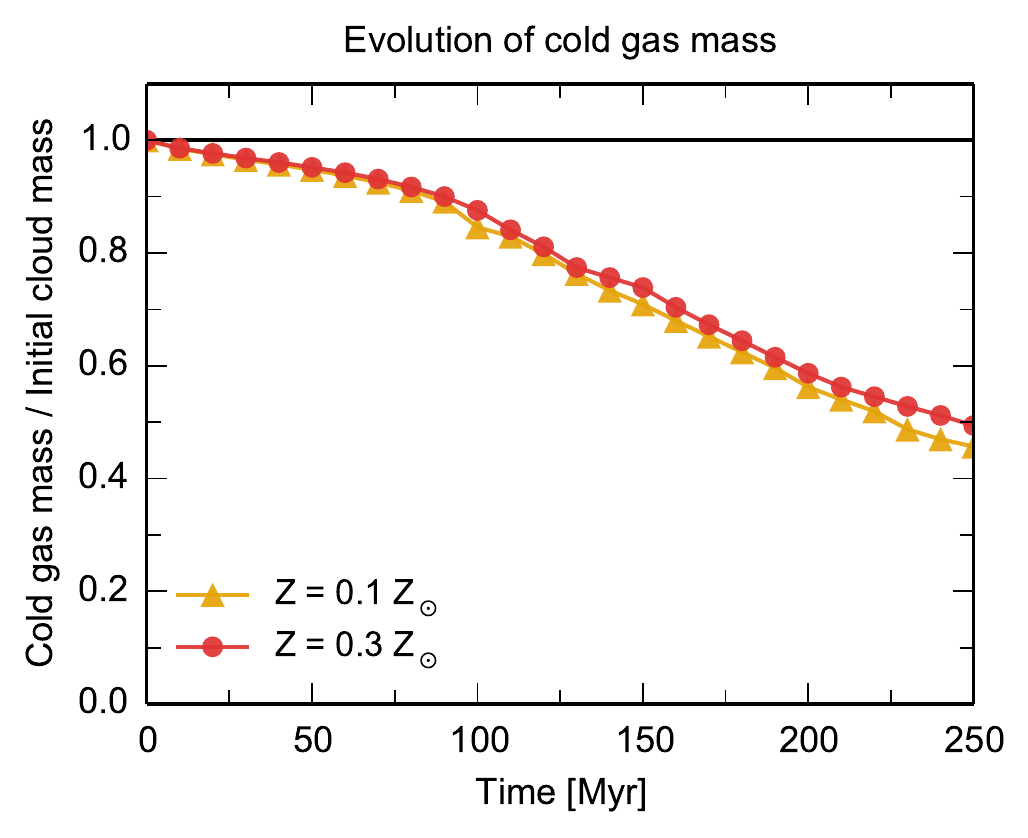}
\caption{Evolution of the mass of cool gas ($T<10^5$ K) with time for two simulations with different cloud metallicity, $Z_\mathrm{cl} = 0.1$ Z$_\odot$ (yellow line) and $Z_\mathrm{cl} = 0.3$ Z$_\odot$ (red line, Sim.7 in Tab.~\ref{ChangedParameters}). Both the simulations starts with initial cloud velocity $100\kms$ and initial cloud radius $250$ pc.}
\label{Fig6}
\end{figure}

All our simulations were performed with cloud metallicity equal to $0.3$ Z$_\odot$. We adopted this value because it is in agreement with the average metallicity measured for most of the high-velocity complexes in the Milky Way (see Sec.~\ref{Hydrodynamical simulations}). So far, the COS-Halos data have not allowed to obtain accurate information about the metallicity of the cool CGM \citep{Werk+14}. In a very recent work, \citet{Prochaska+17} analyzed new far-ultraviolet quasar spectra from the COS-Halos survey and found that the median metallicity of the 32 selected CGM systems is $~ 0.31$ Z$_\odot$, in agreement with the value adopted in this paper.  However, the metallicity distribution spans over a wide range of values, with a $95 \%$ interval in the range from $~1/50$ Z$_\odot$ to 3 Z$_\odot$. 
Different values of gas metallicity could affect the results of our simulations. For instance, lower values of the gas metallicity would imply a decrease of gas cooling rates. In our simulations, reducing the cloud metallicity entails that the metallicity of the gas mixture behind the cloud decreases. Lower cooling rates of the wake could lead to a faster evaporation of the cool cloudlets within it and, therefore, shorter survival times for the cool gas. On the other hand, if the cloud metallicities is higher than $0.3$ Z$_\odot$, the cool gas could survive for a longer time.

In order to evaluate how a lower cloud metallicity can influence the survival of cool gas, we repeated our fiducial simulation (Sim.7 in Tab.~\ref{ChangedParameters}) assuming the same metallicity for cloud and corona, $0.1$ Z$_\odot$. In Fig.~\ref{Fig6} we compare the evolution of the mass of cool gas with time obtained using the two different cloud metallicities.
The amount of cool gas decreases slightly faster in the case at lower metallicity but the difference between the two trends after 250 Myr is lower than $10 \%$. We conclude that our results do not change significantly if we lower the metallicity of the cool gas.

Our only result incompatible with the COS-Halos findings concerns the number gas density. \citet{Werk+14} found that it is very low ($n \sim 10^{-4}$ cm$^{-3}$) and inconsistent with its being in pressure equilibrium with the hot gas phase. This result is very puzzling because it is not clear how the cool material could survive in these conditions. Indeed, we found that clouds initially out of equilibrium reach pressure equilibrium in a few Myrs. Moreover, despite the cool gas density can decrease once the gas is stripped from the cloud, it remains at least an order of magnitude larger than cool gas density found by \citet{Werk+14}. However, this latter remains quite uncertain due to uncertainty in EUVB radiation field used for the modelling of the \hi\ column densities. 

\subsection{Comparison with previous works}
\label{Discussion1}
Simulations of the survival of cool HVCs moving in a hot environment were already performed in the past, showing that the survival timescales are relatively short. \citet{Heitsch&Putman09} found that, for cloud velocities and halo densities typical of the Milky Way, \hi\ clouds are destroyed on timescales of $\sim$ 100 Myr. However, they investigated the survival of \hi\ gas ($T \leq 10^4$ K), while we took into account all gas at $T \leq 10^5$ K, i.e. both neutral and ionized cool gas. Furthermore, they analysed clouds with mass $\lesssim 3\times 10^{4}\, \mo$, while our mass range extends up to $ 1.6\times10^{5} \, \mo$.  In order to better compare their result with ours, we analysed the evolution of gas at $T \leq 10^4$ K in our simulations. We obtained that the mass surviving after 250 Myr is $\sim 35 \%$ of the initial mass for the cloud with $M_\mathrm{cl}=2\times 10^4 \, \mo$ and $v_\mathrm{cl} = 100 \kms$, while it is null for higher velocities. This result seems to agree with those of \citet{Heitsch&Putman09}, accounting that they neglected thermal conduction, which hinders the cloud destruction.
In our simulations with $M_\mathrm{cl}=1.6\times 10^5 \, \mo$ the mass of gas at $T \leq 10^4$ K is less than $50\%$ after 250 Myr.

Short survival timescales (100-200 Myr) were also found by \citet{Bland-Hawthorn+07}. They focused on the origin of $\ha$ emission along the MS and explained it as due to the interaction between the \hi\ Stream and the hot corona: the cool ionized gas is shocked material as it is ablated from the Stream clouds by Kelvin-Helmholtz instabilities at the cloud-corona interface. This result agrees with ours: the large amount of ionized gas at $T\sim10^{4-5}$ K found by \citet{Bland-Hawthorn+07} is comparable to the cool gas mass at $T\leq10^5$ K that survives in our simulations.

Both the simulations above mentioned were performed in a three-dimensional cartesian geometry, while all our simulations are two-dimensional. Indeed, 3D simulations would have imply prohibitive computational times for a code with static grid,
as ATHENA. \citet{Armillotta+16} compared low-resolution simulations of cloud-corona interaction obtained by using the two different geometries. We found that, since in 3D the contact surface between the two fluids is larger, hydrodynamical instabilities and subsequent loss of gas from the cool cloud evolve more quickly. In this work, the effect of 2D geometry could therefore be to slow down the cloud destruction. However, in \citet{Armillotta+16} the difference in terms of mass of cool gas between the two geometries is lower than $10\%$ during the whole computational time, so it does not affect significantly the final result. Moreover, we point out that in the present work thermal conduction is much more important to slow down the cloud destruction than for the simulations of \citet{Armillotta+16}. Thus, it is likely that the larger contact surface in 3D simulations would increase the development of velocity gradients at cloud-corona interface and the efficiency of thermal conduction. 
Therefore, it is unclear whether the combination of thermal conduction and 3D geometry would result in faster or slower cloud destruction, but we do not expect a significant difference with respect to our 2D simulations. 

The simulations in the literature most similar to ours are those of \citet{Kwak+11}, who simulated in a 2D cartesian geometry the motion of HVCs through the Galactic corona. They found that clouds with masses larger than $4\times10^5\, \mo$ remain largely intact after 240 Myr, while we found that also clouds with smaller mass ($M_\mathrm{cl} = 1.6\times10^5\, \mo$) are able to keep a large fraction of their own initial mass ($\gtrsim 70\%$ after 250 Myr). However, as \citet{Heitsch&Putman09}, \citet{Kwak+11} analysed the survival of \hi\ gas, at $T \leq 10^4$ K and neglected thermal conduction.
It is interesting to point out that, similarly to our analysis, they found that high ionization elements, as \ovi, are produced by cloud-corona mixing and lines of sight that cross the turbulent wake detect a large number of high ionization elements but a small numbers of low ionization elements, which are nearly absent in the wake.

\citet{Marinacci+10}, \citet{Fraternali+15} and \citet{Armillotta+16} analysed the cloud motion at the disc-corona interface (few kpc above disc), showing that the cloud is able to trigger the condensation of a large portion of coronal material. This result does not contradict those presented here. Indeed, the coronal density in the region close to the galactic disc is around one order of magnitude larger than the coronal density in the outer halo that we considered here. As a consequence, the density of the gas mixture behind the cloud becomes larger when the cloud is close to the galactic disc. The balance between evaporation and condensation strongly depends on this density, since the cooling rate is proportional to its square. Decreasing the wake density, the gas condensation becomes ineffective and the cool gas stripped from the cloud evaporates.

\section{Conclusions}
\label{Conclusions}
In the last years, sensitive observations have revealed that low-redshift galaxies are embedded in extended haloes of multiphase gas, the circumgalactic medium. An important improvement to our knowledge of the CGM was obtained by the COS-Halos survey, which has detected gas through absorption lines against background QSO spectra for a sample of galaxies at low-redshift, finding that most of the sample galaxies, regardless of their type, are surrounded by large amount of cool and ionized gas ($T \leq 10^5$ K), extending out to impact parameters of 150 kpc from the galaxy centre. 

In this paper we have explored the physical conditions that allow the survival of cool gas in hot galactic coronae. We performed high-resolution hydrodynamical simulations of cool neutral clouds ($T=10^4$ K) travelling through a hot ($T=2\times10^6$ K) and low-density ($n=10^{-4}$ cm$^{-3}$) ambient medium including cooling, heating and thermal conduction. From our simulations, we conclude that the interaction and mixing between cool clouds and a hot corona lead to a gradual loss of gas from the clouds. The survival time of cool gas strongly depends on initial size (mass) of the cloud: clouds with radius $\gtrsim 250$ pc (mass $\gtrsim 2\times10^4 \mo$) are able to keep a large fraction of their own initial mass for hundreds of Myr. 
Thermal conduction appears quite important for the cloud survival since it slows down the cloud destruction.

We compared the column densities of our simulations with the column densities of the COS-Halos observations. The agreement is very good for low-intermediate ionization elements, as \siii\ and \siiii, but not for high ionization elements, as \ovi. However, while \siii\ traces a cool gas phase, mainly located inside the cloud, \ovi\ traces a warmer and widespread gas phase, situated in the turbulent wake behind the cloud, produced by the mixing between the cool gas ablated from the cloud and the hot coronal medium. Accounting that observations can detect multiple clouds along the line of sight, the probability of detecting diffuse and extended wakes is much higher than the probability to intercept compact cloud heads. This fact could explain why the range of \ovi\ column densities predicted by our single-cloud simulations does not overlap with the observations.

We can conclude that our results suggest that the existence and the ubiquity of large amount of cool/warm gas in the CGM is possible because large enough clouds, made compact by the effect of thermal conduction, are able to survive their interaction with the hot corona for several hundreds of Myr. This may have important implications for the gas accretion necessary to fuel star-formation in Milky Way galaxies \citep[see e.g.][]{Fraternali&Tomassetti12}. A recurrent question is, for instance, whether a massive structure like the MS can reach the disc of the Milky Way, feeding the star formation \citep[e.g.,][]{Fox+14}. We can speculate that the cloudlets composing the MS with mass $\sim 10^{5}\, \mo$ can survive the journey through the Galactic corona and provide a significant amount of gas accretion.

\section*{Acknowledgements}
LA acknowledges financial support from MARCO POLO 2015-2016. LA is pleased to thank the University of California Santa Cruz for the hospitality during the first phase of this work. JXP acknowledges partial funding by NASA grants HST-GO-13033.06-A and HST-GO-13846.005-A. LA is grateful to Annalisa Citro for helping her to use CLOUDY and to Yusuke Fujimoto for useful discussions. We acknowledge the CINECA award under the ISCRA initiative, for the availability of high performance computing resources and support. We acknowledge the University of California Santa Cruz to allow us to use the Hyades supercluster to perform some test simulations.

\bibliography{biblio}

\begin{thebibliography}{78}
\expandafter\ifx\csname natexlab\endcsname\relax\def\natexlab#1{#1}\fi

\bibitem[{{Anderson} \& {Bregman}(2010)}]{Anderson&Bregman10}
{Anderson} M.~E., {Bregman} J.~N., 2010, \apj, 714, 320

\bibitem[{{Anderson}, {Churazov} \& {Bregman}(2016){Anderson}, {Churazov}, \&
  {Bregman}}]{Anderson+16}
{Anderson} M.~E., {Churazov} E., {Bregman} J.~N., 2016, \mnras, 455, 227

\bibitem[{{Armillotta}, {Fraternali} \& {Marinacci}(2016){Armillotta},
  {Fraternali}, \& {Marinacci}}]{Armillotta+16}
{Armillotta} L., {Fraternali} F., {Marinacci} F., 2016, \mnras, 462, 4157

\bibitem[{{Balbus}(1986)}]{Balbus86}
{Balbus} S.~A., 1986, \apj, 304, 787

\bibitem[{{Barger}, {Lehner} \& {Howk}(2016){Barger}, {Lehner}, \&
  {Howk}}]{Barger+16}
{Barger} K.~A., {Lehner} N., {Howk} J.~C., 2016, \apj, 817, 91

\bibitem[{{Besla} {et~al}\mbox{.}(2012){Besla}, {Kallivayalil}, {Hernquist},
  {van der Marel}, {Cox}, \& {Kere{\v s}}}]{Besla+12}
{Besla} G., {Kallivayalil} N., {Hernquist} L., {van der Marel} R.~P., {Cox}
  T.~J., {Kere{\v s}} D., 2012, \mnras, 421, 2109

\bibitem[{{Bland-Hawthorn} {et~al}\mbox{.}(2007){Bland-Hawthorn}, {Sutherland},
  {Agertz}, \& {Moore}}]{Bland-Hawthorn+07}
{Bland-Hawthorn} J., {Sutherland} R., {Agertz} O., {Moore} B., 2007, \apjl,
  670, L109

\bibitem[{{Bogd{\'a}n} {et~al}\mbox{.}(2013){Bogd{\'a}n}, {Forman},
  {Vogelsberger}, {Bourdin}, {Sijacki}, {Mazzotta}, {Kraft}, {Jones},
  {Gilfanov}, {Churazov}, \& {David}}]{Bogdan+13}
{Bogd{\'a}n} {\'A}. {et~al.}, 2013, \apj, 772, 97

\bibitem[{{Boomsma} {et~al}\mbox{.}(2008){Boomsma}, {Oosterloo}, {Fraternali},
  {van der Hulst}, \& {Sancisi}}]{Boomsma+08}
{Boomsma} R., {Oosterloo} T.~A., {Fraternali} F., {van der Hulst} J.~M.,
  {Sancisi} R., 2008, \aap, 490, 555

\bibitem[{{Bregman}(2007)}]{Bregman07}
{Bregman} J.~N., 2007, \araa, 45, 221

\bibitem[{{Br{\"u}ns} {et~al}\mbox{.}(2000){Br{\"u}ns}, {Kerp}, {Kalberla}, \&
  {Mebold}}]{Bruens+00}
{Br{\"u}ns} C., {Kerp} J., {Kalberla} P.~M.~W., {Mebold} U., 2000, \aap, 357,
  120

\bibitem[{{Chandran} \& {Cowley}(1998)}]{Chandran&Cowley98}
{Chandran} B.~D.~G., {Cowley} S.~C., 1998, Physical Review Letters, 80, 3077

\bibitem[{{Collins}, {Shull} \& {Giroux}(2007){Collins}, {Shull}, \&
  {Giroux}}]{Collins+07}
{Collins} J.~A., {Shull} J.~M., {Giroux} M.~L., 2007, \apj, 657, 271

\bibitem[{{Cowie} \& {McKee}(1977)}]{Cowie&McKee77}
{Cowie} L.~L., {McKee} C.~F., 1977, \apj, 211, 135

\bibitem[{{Dai} {et~al}\mbox{.}(2012){Dai}, {Anderson}, {Bregman}, \&
  {Miller}}]{Dai+12}
{Dai} X., {Anderson} M.~E., {Bregman} J.~N., {Miller} J.~M., 2012, \apj, 755,
  107

\bibitem[{{Faerman}, {Sternberg} \& {McKee}(2017){Faerman}, {Sternberg}, \&
  {McKee}}]{Faerman+16}
{Faerman} Y., {Sternberg} A., {McKee} C.~F., 2017, \apj, 835, 52

\bibitem[{{Ferland} {et~al}\mbox{.}(2013){Ferland}, {Porter}, {van Hoof},
  {Williams}, {Abel}, {Lykins}, {Shaw}, {Henney}, \& {Stancil}}]{Ferland+13}
{Ferland} G.~J. {et~al.}, 2013, \rmxaa, 49, 137

\bibitem[{{For}, {Staveley-Smith} \& {McClure-Griffiths}(2013){For},
  {Staveley-Smith}, \& {McClure-Griffiths}}]{For+13}
{For} B.-Q., {Staveley-Smith} L., {McClure-Griffiths} N.~M., 2013, \apj, 764,
  74

\bibitem[{{Fox} {et~al}\mbox{.}(2004){Fox}, {Savage}, {Wakker}, {Richter},
  {Sembach}, \& {Tripp}}]{Fox+04}
{Fox} A.~J., {Savage} B.~D., {Wakker} B.~P., {Richter} P., {Sembach} K.~R.,
  {Tripp} T.~M., 2004, \apj, 602, 738

\bibitem[{{Fox} {et~al}\mbox{.}(2014){Fox}, {Wakker}, {Barger}, {Hernandez},
  {Richter}, {Lehner}, {Bland-Hawthorn}, {Charlton}, {Westmeier}, {Thom},
  {Tumlinson}, {Misawa}, {Howk}, {Haffner}, {Ely}, {Rodriguez-Hidalgo}, \&
  {Kumari}}]{Fox+14}
{Fox} A.~J. {et~al.}, 2014, \apj, 787, 147

\bibitem[{{Fox} {et~al}\mbox{.}(2005){Fox}, {Wakker}, {Savage}, {Tripp},
  {Sembach}, \& {Bland-Hawthorn}}]{Fox+05}
{Fox} A.~J., {Wakker} B.~P., {Savage} B.~D., {Tripp} T.~M., {Sembach} K.~R.,
  {Bland-Hawthorn} J., 2005, \apj, 630, 332

\bibitem[{{Fox} {et~al}\mbox{.}(2010){Fox}, {Wakker}, {Smoker}, {Richter},
  {Savage}, \& {Sembach}}]{Fox+10}
{Fox} A.~J., {Wakker} B.~P., {Smoker} J.~V., {Richter} P., {Savage} B.~D.,
  {Sembach} K.~R., 2010, \apj, 718, 1046

\bibitem[{{Fraternali} \& {Binney}(2006)}]{Fraternali&Binney06}
{Fraternali} F., {Binney} J.~J., 2006, \mnras, 366, 449

\bibitem[{{Fraternali} {et~al}\mbox{.}(2015){Fraternali}, {Marasco},
  {Armillotta}, \& {Marinacci}}]{Fraternali+15}
{Fraternali} F., {Marasco} A., {Armillotta} L., {Marinacci} F., 2015, \mnras,
  447, L70

\bibitem[{{Fraternali} \& {Tomassetti}(2012)}]{Fraternali&Tomassetti12}
{Fraternali} F., {Tomassetti} M., 2012, \mnras, 426, 2166

\bibitem[{{Fraternali} {et~al}\mbox{.}(2002){Fraternali}, {van Moorsel},
  {Sancisi}, \& {Oosterloo}}]{Fraternali02}
{Fraternali} F., {van Moorsel} G., {Sancisi} R., {Oosterloo} T., 2002, \aj,
  123, 3124

\bibitem[{{Fukugita} \& {Peebles}(2006)}]{Fukugita&Peebles06}
{Fukugita} M., {Peebles} P.~J.~E., 2006, \apj, 639, 590

\bibitem[{{Gaensler} {et~al}\mbox{.}(2008){Gaensler}, {Madsen}, {Chatterjee},
  \& {Mao}}]{Gaensler+08}
{Gaensler} B.~M., {Madsen} G.~J., {Chatterjee} S., {Mao} S.~A., 2008, \pasa,
  25, 184

\bibitem[{{Gatto} {et~al}\mbox{.}(2013){Gatto}, {Fraternali}, {Read},
  {Marinacci}, {Lux}, \& {Walch}}]{Gatto+13}
{Gatto} A., {Fraternali} F., {Read} J.~I., {Marinacci} F., {Lux} H., {Walch}
  S., 2013, \mnras, 433, 2749

\bibitem[{{Gentile} {et~al}\mbox{.}(2013){Gentile}, {J{\'o}zsa}, {Serra},
  {Heald}, {de Blok}, {Fraternali}, {Patterson}, {Walterbos}, \&
  {Oosterloo}}]{Gentile+13}
{Gentile} G. {et~al.}, 2013, \aap, 554, A125

\bibitem[{{Grcevich} \& {Putman}(2009)}]{Grcevich&Putman09}
{Grcevich} J., {Putman} M.~E., 2009, \apj, 696, 385

\bibitem[{{Haardt} \& {Madau}(2012)}]{Haardt&Madau12}
{Haardt} F., {Madau} P., 2012, \apj, 746, 125

\bibitem[{{Heitsch} \& {Putman}(2009)}]{Heitsch&Putman09}
{Heitsch} F., {Putman} M.~E., 2009, \apj, 698, 1485

\bibitem[{{Hodges-Kluck} \& {Bregman}(2013)}]{Hodges-Kluck&Bregman13}
{Hodges-Kluck} E.~J., {Bregman} J.~N., 2013, \apj, 762, 12

\bibitem[{{Hsu} {et~al}\mbox{.}(2011){Hsu}, {Putman}, {Heitsch},
  {Stanimirovi{\'c}}, {Peek}, \& {Clark}}]{Hsu+11}
{Hsu} W.-H., {Putman} M.~E., {Heitsch} F., {Stanimirovi{\'c}} S., {Peek}
  J.~E.~G., {Clark} S.~E., 2011, \aj, 141, 57

\bibitem[{{Hunter} {et~al}\mbox{.}(2009){Hunter}, {Brott}, {Langer}, {Lennon},
  {Dufton}, {Howarth}, {Ryans}, {Trundle}, {Evans}, {de Koter}, \&
  {Smartt}}]{Hunter+09}
{Hunter} I. {et~al.}, 2009, \aap, 496, 841

\bibitem[{{Kalberla} \& {Haud}(2006)}]{Kalberla+06}
{Kalberla} P.~M.~W., {Haud} U., 2006, \aap, 455, 481

\bibitem[{{Kwak}, {Henley} \& {Shelton}(2011){Kwak}, {Henley}, \&
  {Shelton}}]{Kwak+11}
{Kwak} K., {Henley} D.~B., {Shelton} R.~L., 2011, \apj, 739, 30

\bibitem[{{Lehner} \& {Howk}(2011)}]{Lehner&Howk+11}
{Lehner} N., {Howk} J.~C., 2011, Science, 334, 955

\bibitem[{{Lehner} {et~al}\mbox{.}(2012){Lehner}, {Howk}, {Thom}, {Fox},
  {Tumlinson}, {Tripp}, \& {Meiring}}]{Lehner+12}
{Lehner} N., {Howk} J.~C., {Thom} C., {Fox} A.~J., {Tumlinson} J., {Tripp}
  T.~M., {Meiring} J.~D., 2012, \mnras, 424, 2896

\bibitem[{{Lewis} {et~al}\mbox{.}(2013){Lewis}, {Braun}, {McConnachie},
  {Irwin}, {Ibata}, {Chapman}, {Ferguson}, {Martin}, {Fardal}, {Dubinski},
  {Widrow}, {Mackey}, {Babul}, {Tanvir}, \& {Rich}}]{Lewis+13}
{Lewis} G.~F. {et~al.}, 2013, \apj, 763, 4

\bibitem[{{Lux} {et~al}\mbox{.}(2013){Lux}, {Read}, {Lake}, \&
  {Johnston}}]{Lux+13}
{Lux} H., {Read} J.~I., {Lake} G., {Johnston} K.~V., 2013, \mnras, 436, 2386

\bibitem[{{Marasco} \& {Fraternali}(2017)}]{Marasco+17}
{Marasco} A., {Fraternali} F., 2017, \mnras, 464, L100

\bibitem[{{Marcolini} {et~al}\mbox{.}(2005){Marcolini}, {Strickland},
  {D'Ercole}, {Heckman}, \& {Hoopes}}]{Marcolini+05}
{Marcolini} A., {Strickland} D.~K., {D'Ercole} A., {Heckman} T.~M., {Hoopes}
  C.~G., 2005, \mnras, 362, 626

\bibitem[{{Marinacci} {et~al}\mbox{.}(2010){Marinacci}, {Binney}, {Fraternali},
  {Nipoti}, {Ciotti}, \& {Londrillo}}]{Marinacci+10}
{Marinacci} F., {Binney} J., {Fraternali} F., {Nipoti} C., {Ciotti} L.,
  {Londrillo} P., 2010, \mnras, 404, 1464

\bibitem[{{Mastropietro} {et~al}\mbox{.}(2005){Mastropietro}, {Moore}, {Mayer},
  {Wadsley}, \& {Stadel}}]{Mastropietro+05}
{Mastropietro} C., {Moore} B., {Mayer} L., {Wadsley} J., {Stadel} J., 2005,
  \mnras, 363, 509

\bibitem[{{Miller} \& {Bregman}(2015)}]{Miller&Bregman15}
{Miller} M.~J., {Bregman} J.~N., 2015, \apj, 800, 14

\bibitem[{{Murray} {et~al}\mbox{.}(1993){Murray}, {White}, {Blondin}, \&
  {Lin}}]{Murray+93}
{Murray} S.~D., {White} S.~D.~M., {Blondin} J.~M., {Lin} D.~N.~C., 1993, \apj,
  407, 588

\bibitem[{{Narayan} \& {Medvedev}(2001)}]{Narayan&Medvedev01}
{Narayan} R., {Medvedev} M.~V., 2001, \apjl, 562, L129

\bibitem[{{Oosterloo}, {Fraternali} \& {Sancisi}(2007){Oosterloo},
  {Fraternali}, \& {Sancisi}}]{Oosterloo+07}
{Oosterloo} T., {Fraternali} F., {Sancisi} R., 2007, \aj, 134, 1019

\bibitem[{{Prochaska} {et~al}\mbox{.}(2017){Prochaska}, {Werk}, {Worseck},
  {Tripp}, {Tumlinson}, {Burchett}, {Fox}, {Fumagalli}, {Lehner}, {Peeples}, \&
  {Tejos}}]{Prochaska+17}
{Prochaska} J.~X. {et~al.}, 2017, ArXiv e-prints

\bibitem[{{Putman}, {Peek} \& {Joung}(2012){Putman}, {Peek}, \&
  {Joung}}]{Putman+12}
{Putman} M.~E., {Peek} J.~E.~G., {Joung} M.~R., 2012, \araa, 50, 491

\bibitem[{{Putman}, {Saul} \& {Mets}(2011){Putman}, {Saul}, \&
  {Mets}}]{Putman+11}
{Putman} M.~E., {Saul} D.~R., {Mets} E., 2011, \mnras, 418, 1575

\bibitem[{{Putman} {et~al}\mbox{.}(2003){Putman}, {Staveley-Smith}, {Freeman},
  {Gibson}, \& {Barnes}}]{Putman+03}
{Putman} M.~E., {Staveley-Smith} L., {Freeman} K.~C., {Gibson} B.~K., {Barnes}
  D.~G., 2003, \apj, 586, 170

\bibitem[{{Salem} {et~al}\mbox{.}(2015){Salem}, {Besla}, {Bryan}, {Putman},
  {van der Marel}, \& {Tonnesen}}]{Salem+15}
{Salem} M., {Besla} G., {Bryan} G., {Putman} M., {van der Marel} R.~P.,
  {Tonnesen} S., 2015, \apj, 815, 77

\bibitem[{{Sancisi} {et~al}\mbox{.}(2008){Sancisi}, {Fraternali}, {Oosterloo},
  \& {van der Hulst}}]{Sancisi+08}
{Sancisi} R., {Fraternali} F., {Oosterloo} T., {van der Hulst} T., 2008, \aapr,
  15, 189

\bibitem[{{Savage} {et~al}\mbox{.}(2003){Savage}, {Sembach}, {Wakker},
  {Richter}, {Meade}, {Jenkins}, {Shull}, {Moos}, \& {Sonneborn}}]{Savage+03}
{Savage} B.~D. {et~al.}, 2003, \apjs, 146, 125

\bibitem[{{Scannapieco} \& {Br{\"u}ggen}(2015)}]{Scannapieco&Bruggen15}
{Scannapieco} E., {Br{\"u}ggen} M., 2015, \apj, 805, 158

\bibitem[{{Sembach} {et~al}\mbox{.}(2003){Sembach}, {Wakker}, {Savage},
  {Richter}, {Meade}, {Shull}, {Jenkins}, {Sonneborn}, \& {Moos}}]{Sembach+03}
{Sembach} K.~R. {et~al.}, 2003, \apjs, 146, 165

\bibitem[{{Shull} {et~al}\mbox{.}(2009){Shull}, {Jones}, {Danforth}, \&
  {Collins}}]{Shull+09}
{Shull} J.~M., {Jones} J.~R., {Danforth} C.~W., {Collins} J.~A., 2009, \apj,
  699, 754

\bibitem[{{Shull} {et~al}\mbox{.}(2011){Shull}, {Stevans}, {Danforth},
  {Penton}, {Lockman}, \& {Arav}}]{Shull+11}
{Shull} J.~M., {Stevans} M., {Danforth} C., {Penton} S.~V., {Lockman} F.~J.,
  {Arav} N., 2011, \apj, 739, 105

\bibitem[{{Spitzer}(1956)}]{Spitzer56}
{Spitzer} L., 1956, {Physics of Fully Ionized Gases}

\bibitem[{{Spitzer}(1962)}]{Spitzer62}
{Spitzer} L., 1962, {Physics of Fully Ionized Gases}

\bibitem[{{Stone} {et~al}\mbox{.}(2008){Stone}, {Gardiner}, {Teuben}, {Hawley},
  \& {Simon}}]{Stone+08}
{Stone} J.~M., {Gardiner} T.~A., {Teuben} P., {Hawley} J.~F., {Simon} J.~B.,
  2008, \apjs, 178, 137

\bibitem[{{Sutherland} \& {Dopita}(1993)}]{Sutherland&Dopita93}
{Sutherland} R.~S., {Dopita} M.~A., 1993, \apjs, 88, 253

\bibitem[{{Thilker} {et~al}\mbox{.}(2004){Thilker}, {Braun}, {Walterbos},
  {Corbelli}, {Lockman}, {Murphy}, \& {Maddalena}}]{Thilker+04}
{Thilker} D.~A., {Braun} R., {Walterbos} R.~A.~M., {Corbelli} E., {Lockman}
  F.~J., {Murphy} E., {Maddalena} R., 2004, \apjl, 601, L39

\bibitem[{{Thom} {et~al}\mbox{.}(2008){Thom}, {Peek}, {Putman}, {Heiles},
  {Peek}, \& {Wilhelm}}]{Thom+08}
{Thom} C., {Peek} J.~E.~G., {Putman} M.~E., {Heiles} C., {Peek} K.~M.~G.,
  {Wilhelm} R., 2008, \apj, 684, 364

\bibitem[{{Tumlinson} {et~al}\mbox{.}(2013){Tumlinson}, {Thom}, {Werk},
  {Prochaska}, {Tripp}, {Katz}, {Dav{\'e}}, {Oppenheimer}, {Meiring}, {Ford},
  {O'Meara}, {Peeples}, {Sembach}, \& {Weinberg}}]{Tumlinson+13}
{Tumlinson} J. {et~al.}, 2013, \apj, 777, 59

\bibitem[{{Tumlinson} {et~al}\mbox{.}(2011){Tumlinson}, {Thom}, {Werk},
  {Prochaska}, {Tripp}, {Weinberg}, {Peeples}, {O'Meara}, {Oppenheimer},
  {Meiring}, {Katz}, {Dav{\'e}}, {Ford}, \& {Sembach}}]{Tumlinson+11}
{Tumlinson} J. {et~al.}, 2011, Science, 334, 948

\bibitem[{{Venzmer}, {Kerp} \& {Kalberla}(2012){Venzmer}, {Kerp}, \&
  {Kalberla}}]{Venzmer+12}
{Venzmer} M.~S., {Kerp} J., {Kalberla} P.~M.~W., 2012, \aap, 547, A12

\bibitem[{{Vieser} \& {Hensler}(2007)}]{Vieser&Hensler07b}
{Vieser} W., {Hensler} G., 2007, \aap, 472, 141

\bibitem[{{Wakker}(2001)}]{Wakker01}
{Wakker} B.~P., 2001, \apjs, 136, 463

\bibitem[{{Wakker} \& {van Woerden}(1997)}]{Wakker&vanWoerden97}
{Wakker} B.~P., {van Woerden} H., 1997, \araa, 35, 217

\bibitem[{{Wakker} {et~al}\mbox{.}(2007){Wakker}, {York}, {Howk}, {Barentine},
  {Wilhelm}, {Peletier}, {van Woerden}, {Beers}, {Ivezi{\'c}}, {Richter}, \&
  {Schwarz}}]{Wakker07}
{Wakker} B.~P. {et~al.}, 2007, \apjl, 670, L113

\bibitem[{{Werk} {et~al}\mbox{.}(2016){Werk}, {Prochaska}, {Cantalupo}, {Fox},
  {Oppenheimer}, {Tumlinson}, {Tripp}, {Lehner}, \& {McQuinn}}]{Werk+16}
{Werk} J.~K. {et~al.}, 2016, ArXiv e-prints

\bibitem[{{Werk} {et~al}\mbox{.}(2012){Werk}, {Prochaska}, {Thom}, {Tumlinson},
  {Tripp}, {O'Meara}, \& {Meiring}}]{Werk+12}
{Werk} J.~K., {Prochaska} J.~X., {Thom} C., {Tumlinson} J., {Tripp} T.~M.,
  {O'Meara} J.~M., {Meiring} J.~D., 2012, \apjs, 198, 3

\bibitem[{{Werk} {et~al}\mbox{.}(2013){Werk}, {Prochaska}, {Thom}, {Tumlinson},
  {Tripp}, {O'Meara}, \& {Peeples}}]{Werk+13}
{Werk} J.~K., {Prochaska} J.~X., {Thom} C., {Tumlinson} J., {Tripp} T.~M.,
  {O'Meara} J.~M., {Peeples} M.~S., 2013, \apjs, 204, 17

\bibitem[{{Werk} {et~al}\mbox{.}(2014){Werk}, {Prochaska}, {Tumlinson},
  {Peeples}, {Tripp}, {Fox}, {Lehner}, {Thom}, {O'Meara}, {Ford}, {Bordoloi},
  {Katz}, {Tejos}, {Oppenheimer}, {Dav{\'e}}, \& {Weinberg}}]{Werk+14}
{Werk} J.~K. {et~al.}, 2014, \apj, 792, 8

\end{thebibliography}
\label{lastpage}

\end{document}